\documentclass[%
 aps,prb,twocolumn,superscriptaddress,
 amsmath,amssymb,amsfonts,
 floatfix
]{revtex4-2}
\usepackage[T1]{fontenc}
\usepackage{lmodern}
\usepackage{microtype}
\usepackage{graphicx}
\usepackage{bm}
\usepackage{appendix}
\usepackage[dvipsnames]{xcolor}
\usepackage{hyperref}
\hypersetup{
    colorlinks=true,
    citecolor=Green,
    linkcolor=Red,
    urlcolor=NavyBlue
}
\usepackage[capitalise]{cleveref}
\usepackage{soul}
\usepackage{mathtools}
\usepackage{siunitx}

\renewcommand{\vec}[1]{\mathbf{#1}}

\DeclarePairedDelimiter\abs{\lvert}{\rvert}
\DeclarePairedDelimiter\expval{\langle}{\rangle}

\DeclarePairedDelimiter\ket{\lvert}{\rangle}

\begin{abstract}
Quantum geometry quantifies how the single-particle Bloch wavefunction changes in phase and amplitude across the Brillouin Zone.
In multi-orbital systems where bands have strongly mixed orbital composition, quantum geometry plays a vital role in determining the ground state and low-energy properties of interacting electronic systems. 
In this work, we show that Mott metal-insulator transitions, as well as transitions between different magnetic orders within the Mott insulating phase, can be driven by the quantum geometry of the underlying Bloch band, thereby providing a mechanism complementary to conventional bandwidth-tuned Mott transitions.
By studying the Kane-Mele-Hubbard model using exact diagonalization, we demonstrate that in in half-filled and topologically-trivial bands, quantum geometric properties of the Bloch states alone can act as a tuning knob for Mott metal-to-insulator and affect the competition between ferromagnetism and antiferromagnetism. We show that both transitions may be heuristically understood via non-local Coulomb scattering in a basis of exponentially localized Wannier functions. These results highlight the role of quantum geometry beyond topological settings as a governing principle for conventional Mott and magnetic physics in multi-orbital and moir\'e materials.
\end{abstract}

\begin{document} 

\title{Quantum-geometry-driven Mott transitions and magnetism}

\author{Jixun K. Ding}
\email{jkding@sas.upenn.edu}
\affiliation{Department of Physics and Astronomy, University of Pennsylvania, Philadelphia, Pennsylvania 19104, USA}

\author{Martin Claassen}
\email{claassen@sas.upenn.edu}
\affiliation{Department of Physics and Astronomy, University of Pennsylvania, Philadelphia, Pennsylvania 19104, USA}

\date{\today}

\maketitle 

\section{Introduction} 
\label{sec:intro}

Quantum geometry has recently emerged as an essential ingredient in characterizing and designing quantum materials~\cite{Torma2023}. While its impact on the ground-state and dynamical properties of non-interacting and weakly interacting systems is now well established~\cite{Xiao2010,Gao2025c}, our understanding of how Bloch band quantum geometry influences the ground state phase diagram of strongly interacting particles remains at its infancy. 
Some strides in this direction include the notion that band ideality~\cite{Roy2014,Claassen2015,Ledwith2023} favors the formation of fractional quantum anomalous Hall states. Quantum geometry (in the form of integrated quantum metric) has also been shown to provide a boost to superconductivity~\cite{Torma2022,Herzog-Arbeitman2022a,Tian2023,Zeng2025}, superfluidity~\cite{Verma2024} and ferromagnetism~\cite{Chen2025e,Wu2020b,Kang2024b}. 
``Quantum geometric nesting'' conditions~\cite{Han2024,Zhang2025} have been identified to determine interacting parent Hamiltonians for charge, pairing, or spin orders. However, how quantum geometry influences the formation of Mott insulators in topologically trivial bands -- arguably the paradigmatic example of strongly correlated matter -- has not yet been studied. This is an important question to address, particularly in the context of two-dimensional moir\'e materials~\cite{Chen2019a,Tang2020,Ghiotto2021,Li2021b}, where band-geometric effects are believed to be pronounced. 

In this work, we isolate and highlight the underappreciated role of quantum geometry as a driver for Mott metal-to-insulator transitions (MIT).
In the standard single-band Hubbard model, a bandwidth-tuned Mott transition occurs when strong local Coulomb repulsion exceeds the single-particle bandwidth and causes the formation of local moments, which sit on perfectly localized Wannier orbitals on each lattice site~\cite{Imada1998}. However, the presence of quantum geometry sets a lower bound on Wannier localization and forces orbitals to have finite extent~\cite{Marzari2012}. These ``imperfectly localized'' orbitals should be expected to lead to a reduced energy penalty for moving charges, both via lowering the effective on-site repulsion and via enabling density-assisted hopping due to nonzero overlap with their neighbors. Thus, increasing the minimum orbital spread, even while keeping bare bandwidth and interactions fixed, drives a Mott insulator towards a metal. We demonstrate that this is indeed the case by performing band-projected exact diagonalization (ED) calculations on the quarter-filled interacting Kane-Mele~\cite{Kane2005,Kane2005a} model.
We further analyze the magnetic ground states of the Mott insulating phase, and corroborate recent proposals that quantum geometry contributes to ferromagnetic–antiferromagnetic (FM-AFM) transitions in nearly flat bands~\cite{Repellin2020a,Hu2025a,Oh2025a}. 
Both the Mott transition and the magnetic phase transition can be understood by examining short-range projected interaction elements in a basis of exponentially localized Wannier functions.

Our study demonstrates that in addition to the better known bandwidth and filling driven Mott transitions~\cite{Imada1998}, there exists a distinct mechanism of ``quantum-geometric Mott transitions'' tuned by the structure of Bloch wavefunctions that constitute the partially occupied band. Such a mechanism may play a role in Mott transitions observed in Ref.~\cite{Chen2019a,Li2021b,Ghiotto2021}, and may be observable in other transition metal dichalcogenide (TMD) moir\'e bilayer structures where the topmost moir\'e valence bands do not have charge Chern numbers, such as tMoTe$_2$ at small twist angles~\cite{Zhang2024h}. We stress that our theory applies to Mott insulators in unobstructed, topologically trivial bands, which distinguishes our scenario from topologically nontrivial band structures that generically exihibit flavor ferromagnetism and the quantum anomalous Hall effect at integer filling factors~\cite{Yu2014,Devakul2022,Mai2023c,Ledwith2025a}.

\section{Model} 
\label{sec:methods}

We start with a multi-orbital tight-binding model of spinful electrons, where due to orbital mixing, each individual band has non-trivial quantum geometrical information, captured either by the quantum geometric tensor~\cite{Provost1980}, or some generalization thereof. We assume one isolated band is nearly flat, with bandwidth $W$, and electrons interact with each other via a bare Hubbard interaction $U$. Upon half-filling the isolated (topologically trivial but geometrically rich) band, we study a regime of energy scales where both the bandwidth of the partially filled band and Hubbard interaction are much smaller than the band gap to other bands. This energy hierarchy is analogous to studies of fractional Chern insulators and allows studying the half-filled band in isolation and examining its interacting ground state properties via band-projected ED calculations, while the influence of remote bands is accounted for via the momentum-space structure of Bloch wavefunctions.
`
Specifically, we study time-reversal-symmetric two-band models of spinful fermions of the form 
\begin{equation}
    \hat{H} = \hat{H}_0 + \hat{V} \label{eq:hv-model}
\end{equation}
where the interaction term $\hat{V}$ is a purely local, intra-orbital Hubbard interaction term 
\begin{equation}
    \hat{V} = \frac{U}{N}\sum_{\substack{\vec{q} \vec{k} \vec{k}' \\\alpha}} c_{\alpha \vec{k} +\vec{q}\uparrow}^\dagger c_{\alpha \vec{k}'-\vec{q}\downarrow}^\dagger c_{\alpha\vec{k}'\downarrow} c_{\alpha\vec{k}\uparrow}, \label{eq:local-interactions}
\end{equation}
and the kinetic term $\hat{H}_0$ may be written as $\hat{H}_0 = \sum_\vec{k} c^\dagger_{\alpha\vec{k}s} \left[h_s(\vec{k})\right]_{\alpha\beta} c_{\beta\vec{k}s}$. In the above, $N$ is the total number of unit cells, $c_{\alpha\vec{k}s}^\dagger$ creates a fermion in orbital $\alpha \in \{A, B\}$ with momentum $\vec{k}$ and spin $s \in \{\uparrow, \downarrow\}$ .
$\hat{H}_0$ in this (momentum, orbital) basis has matrix elements that can be written in term of Pauli matrices $\bm{\sigma} = \left(\sigma_x,\sigma_y,\sigma_z\right)$:
\begin{multline}
h_\uparrow(\vec{k}) = \vec{a}(\vec{k}) \cdot \bm{\sigma} = \begin{bmatrix}
    a_3(\vec{k}) & a_1(\vec{k}) - i a_2(\vec{k}) \\
    a_1(\vec{k}) + ia_2(\vec{k}) & -a_3(\vec{k})
\end{bmatrix} \\
h_\downarrow(\vec{k}) = h^*_\uparrow(-\vec{k}), \label{eq:hup-hdn}
\end{multline}
which has dispersions $\epsilon(\vec{k}) = \pm |\vec{a}(\vec{k})|$ and normalized eigenvectors $\ket{u^s_n(\vec{k})}$, $n \in \{0,1\}$. 
A spinful model constructed as \cref{eq:hup-hdn} is time-reversal symmetric and has $\mathbb{Z}_2 \times U(1)$ spin symmetry. 

To study the effect of bandwidth-driven and geometry-driven Mott transitions in isolation, we consider a deformed single-particle bandstructure: We first flatten the non-interacting bands by instead considering a modified kinetic Hamiltonian $\hat{H}_0 = \sum_\vec{k} c^\dagger_{\alpha\vec{k}s} \left[\tilde{h}_s(\vec{k})\right]_{\alpha\beta} c_{\beta\vec{k}s}$, where $\tilde{h}_s(\vec{k}) = h_s(\vec{k}) / |\epsilon(\vec{k})|$, resulting in two flat bands at energy $\pm 1$ which retain the same Bloch states as the original tight-binding model. This band flattening procedure is equivalent to introducing long-range, exponentially decaying hopping elements in real space for the bare two-band model~\cite{Neupert2011,Parameswaran2013}, so the deformed model still retains spatial locality. Then, we add by hand a small nearest neighbor hopping term $t_{W}$ between adjacent Wannier orbitals of the lower band, 
inducing a small spin-independent dispersion $\varepsilon(\vec{k}) \propto t_{W} \propto W$, which does not change the eigenvectors $\ket{u^s_n(\vec{k})}$. The exact expression for $\varepsilon(\vec{k})$ depends on lattice geometry and will be specified in \cref{sec:results}.

We focus on overall quarter filling (half filling the lower band). When the lower band is partially filled, if interactions are weak enough such that real interband excitations are suppressed, then all the low-energy physics should take place in only a single band, but will be sensitive to the orbital composition of this band. This is where quantum geometry enters -- 
by tuning the tight-binding model parameters to approach (but not cross) a topological phase transition, we change the orbital composition of the lower band, and study its effect on the ground state properties of the interacting system.

To lowest order, quantum geometric effects can be treated by truncating the Hilbert space of \cref{eq:hv-model} to only states within the lower band and solve the projected Hamiltonian within the restricted Hilbert space. To do so, we rewrite the Hamiltonian in terms of band creation and annihilation operators $b^\dagger/b$, which are related to orbital creation and annihilation operators via orbital components of eigenvectors $\ket{u_n^s(\vec{k})}$,
\begin{equation}
    c_{\vec{k}\alpha s} = \sum_{n} u^s_{\alpha n \vec{k}} b_{n\vec{k} s}, \ u^s_{\alpha n \vec{k}} = \langle \alpha\vec{k} \ket{u^s_n(\vec{k})} \label{eq:uank}.
\end{equation}
Thus defined, $b_{n\vec{k}s}^\dagger$ creates a fermion in band $n$ with momentum $\vec{k}$ and spin $s$. In this basis, the projected Hamiltonian with both kinetic and interaction terms is 
\begin{multline}
    \hat{H}' = \hat{H}_{0,\mathrm{eff}} + \hat{V}_{\mathrm{eff}} = \sum_{\vec{k}}\varepsilon(\vec{k}) b_{\vec{k}s}^\dagger b_{\vec{k}s} + \\ \frac{U}{N}
    \sum_{\substack{\vec{q} \vec{k} \vec{k}' \\\alpha}}  
    u^{\uparrow *}_{\alpha \vec{k} + \vec{q}}
    u^{\downarrow *}_{\alpha \vec{k}' - \vec{q}}
    u^{\downarrow}_{\alpha \vec{k}'}
    u^{\uparrow}_{\alpha \vec{k}} 
    b_{\vec{k} + \vec{q}\uparrow}^\dagger 
    b_{\vec{k}' - \vec{q}\downarrow}^\dagger 
    b_{\vec{k}'\downarrow} 
    b_{\vec{k}\uparrow}, \label{eq:proj-final-momentum}
\end{multline}
where we have omitted the band index $n=0$. This is the model we solve using exact diagonalization (ED) in momentum basis, while
controlling the effects of quantum geometry (via projected interaction form factors) and dispersion (via $t_W$) separately.
Projected to the half-filled band, there are only two scales left in the problem: the energy scale $U/t_W$, and a dimensionless scale that alters the momentum-space structure of Bloch wavefunctions. In subsequent discussions, we fix $U=1$, and omit units for $t_W$, ground state energy, and charge gap, with the understanding that they are expressed in terms of $U$.

We first consider two simple limits of \cref{eq:proj-final-momentum}.
The first is the flat band limit, $t_W = 0$. In this case, \cref{eq:proj-final-momentum} has ferromagnetic ground states. A fully polarized Slater determinant $\ket{\mathrm{FM}} = \prod_{\vec{k}} b^\dagger_{\vec{k}\uparrow}\ket{\varnothing}$ is annihilated by the interaction term; since $\hat{V}_{\mathrm{eff}}$ is positive semidefinite, zero energy eigenstates are ground states~\cite{Zhang2019,Abouelkomsan2025}. At the same time, electrons in a half-filled, topologically trivial, dispersionless band with repulsive interactions are expected to form a Mott insulator. Thus, at $t_W = 0$ we expect to observe a ferromagnetic Mott insulator (Mott FM)~\cite{Zhang2019}, equivalent to the ``flat band ferromagnet'' studied by Tasaki and Mielke~\cite{Tasaki2020}. 
The second limit is the ``trivial quantum geometry'' limit, which corresponds, without loss of generality, to choosing form factors $u_{A\vec{k}}^\sigma = 1$ and $u_{B\vec{k}}^\sigma = 0$  for all $\vec{k},\sigma$. This yields isotropic scattering in momentum space involving a single orbital, and reduces the problem to a single-band Hubbard model with hopping $t_W$ and local repulsion $U$. This model is generically a paramagntic metal at small $U/t_W$ and an antiferromagnetic Mott insulator (Mott AFM) at large $U/t_W$. Thus, for finite $t_W$ and non-trivial form factors in \cref{eq:proj-final-momentum}, we expect to observe transitions between Mott FM, Mott AFM and metallic ground states.

For concrete numerical study, we choose the kinetic hamiltonian $\hat{H}_0$ to be the Kane-Mele model~\cite{Kane2005a,Kane2005}.
The Kane-Mele model is a canonical example of a two-band model which can be tuned smoothly though a band inversion, and is either a trivial band insulator or a quantum spin Hall insulator at overall half-filling. It has been argued to capture essential features of the topmost valence bands of twisted MoTe$_2$ bilayers~\cite{Wu2019} and MoTe$_2$/WSe$_2$ heterobilayers~\cite{Luo2023b}. It has been studied in the presence of Hubbard interactions, but primarily at overall half-filling~\cite{Hohenadler2011}. Moreover, extant studies generally consider the case when the single-particle band structure is topological, and when Hubbard interaction $U$ is greater than band gap $\Delta$~\cite{Rachel2010,Hohenadler2011,Zheng2011,Hickey2016,Mai2023c,He2024a}, which is a regime with both single-particle band topology and strong band-mixing. The partially-filled topological band cannot form a Mott insulator with local moments~\cite{Ledwith2025a}, and when band mixing is strong, single-band quantum geometry is no longer a useful description. As such, the topologically trivial, $U,W\ll \Delta$ regime we consider in this work is quite distinct from existing studies. 

\section{Results} 
\label{sec:results}

The Kane-Mele model~\cite{Kane2005,Kane2005a} without Rashba spin-orbit coupling consists of two time-reversed copies of the Haldane model~\cite{Haldane1988}, which has 
\begin{equation}
\vec{a}(\vec{k}) = \begin{bmatrix}
        -t \sum_{\bm{\delta}}\cos(\vec{k}\cdot\bm{\delta}) \\ 
        -t \sum_{\bm{\delta}}\sin(\vec{k}\cdot\bm{\delta}) \\
        -M+2t'\sum_{\bm{\delta}'}\sin(\vec{k}\cdot\bm{\delta}')
    \end{bmatrix} \label{eq:haldane-params}
\end{equation} 
where we assumed that each next-nearest neighbor hopping picks up only $i$ or $-i$ phase. The $\bm{\delta},\bm{\delta}'$ vectors are defined using a convention where real space primitive lattice vectors are $\vec{a}_1 = a(1,0)$ and $\vec{a}_2 = a/2(1,\sqrt{3})$. $\bm{\delta} \in \{ \bm{\delta}_1,\bm{\delta}_2,\bm{\delta}_3 \}$ connects an atom in sublattice $A(B)$ to its three nearest neighbors $B(A)$ atoms located at
$\bm{\delta}_{1,2} = a/2(\pm 1,\sqrt{3}), \bm{\delta}_3 = a(0,-1/\sqrt{3})$. $\bm{\delta}'\in \{ \bm{\delta}'_1,\bm{\delta}'_2,\bm{\delta}'_3 \}$ connects an atom in sublattice $A(B)$ to three out of six of its next-nearest neighbor $A(B)$ atoms located at $\bm{\delta}'_1 =  \vec{a}_1 , \bm{\delta}'_2 = -\vec{a}_2 , \bm{\delta}'_3 = \vec{a}_2 - \vec{a}_1, \bm{\delta}'_4 = -\bm{\delta}'_1, \bm{\delta}'_5 = -\bm{\delta}'_2, \bm{\delta}'_6 = -\bm{\delta}'_3$.

\begin{figure}[htbp]
    \centering
    \includegraphics[width=0.48\textwidth]{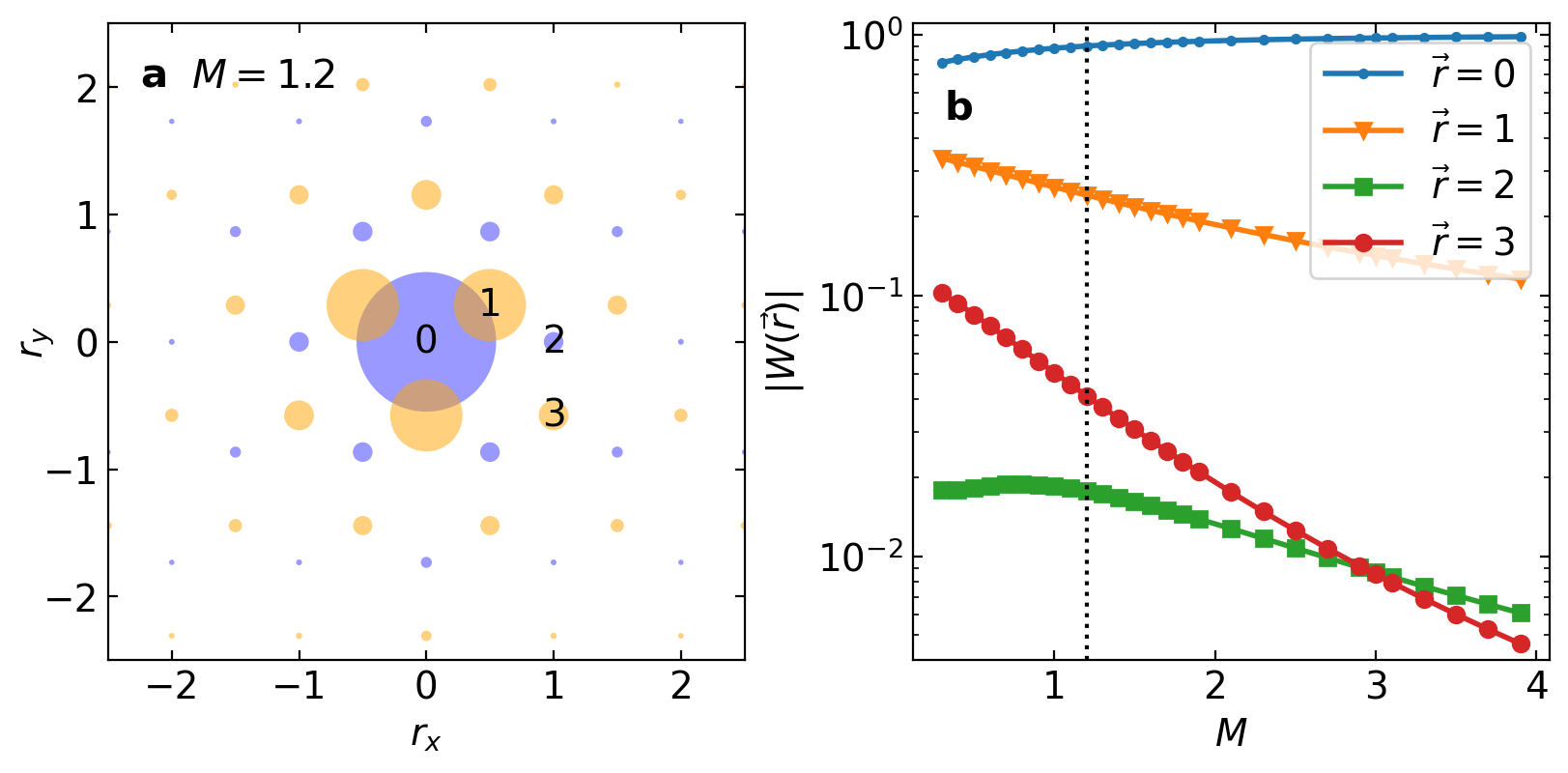}
    \caption{Wannier function evolution of the Kane-Mele model at fixed $t'=0.05$. \textbf{a} Representative real space spread of the Wannier function of the lower band of the Kane-Mele model at $M=1.2$. The size of the markers is proportional to the amplitude $\abs{W_{\vec{R}=\mathbf{0}}(\vec{r})}$ on the original orbitals of the hexagonal lattice, with purple discs for orbital A and orange discs for orbital B. \textbf{b} $M$ dependence of Wannier function amplitude $\abs{W_{\vec{R}=\mathbf{0}}(\vec{r})}$ at four $\vec{r}$ points annotated in \textbf{a} in order of increasing distance from the origin. The black dotted line marks $M = 1.2$.}
    \label{fig:km-Wannier}
\end{figure}

We fix $t=1$ and $|t'/t| < 1/3$, so the Kane-Mele model only has one remaining parameter: the sublattice imbalance $M$. The model has two topological bands with spin Chern number $C_s=\pm 2$ when $|M|< 3\sqrt{3} |t'|$ and two trivial bands with $C_s=0$ if $|M| > 3\sqrt{3} |t'|$. 
We focus on the topologically trivial case with $t' = 0.05$, $M > 3\sqrt{3} t' = 0.26$, and solve \cref{eq:proj-final-momentum} with added dispersion 
\begin{equation}
    \varepsilon(\vec{k}) = -2t_W \sum_{\bm{\delta}'} \cos \vec{k} \cdot \bm{\delta}'.
\end{equation}
In the interacting model, while $t_W$ controls the dispersion, the sublattice mass parameter $M$ controls quantum geometry. This can be seen in \cref{fig:km-Wannier}, where Wannier functions of the lower band and their $M$ dependence are visualized. A smaller value of $M$ means larger orbital mixing, more extended Wannier functions, and less localized charges and magnetic moments in each band.

\begin{figure*}[htbp]
    \centering
    \includegraphics[width=\textwidth]{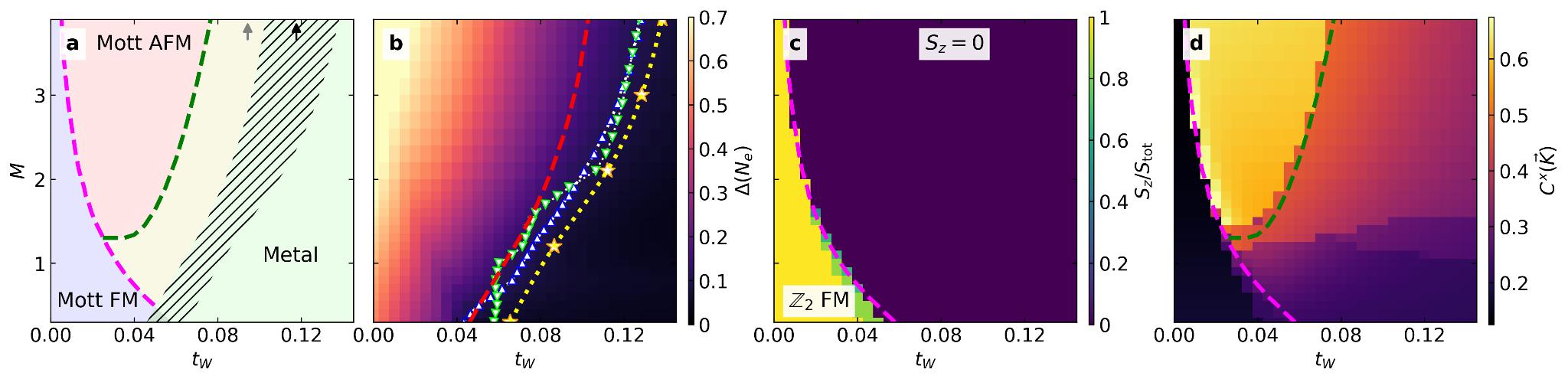}
    \caption{Quarter-filled Kane-Mele model results. \textbf{a} Schematic phase diagram of charge and spin order, as a function of bare dispersion $t_W$ and orbital imbalance $M$. Magenta line dashed denotes the FM-AFM magnetic transition. Green dashed line denotes the possible location of an AFM - remnant region (described in the main text) transition. Hatched region denotes approximate location of Mott transition. The hatched region is identical to the area between red dashed and yellow dotted line in \textbf{b}. The black and grey arrows indicate the position of the metal-to-insulator and magnetic ordering transitions, respectively, of the triangular lattice Hubbard model~\cite{Szasz2020}. \textbf{b} Ground state charge gap $\Delta(N_e)$. Blue triangles, bright green triangles, yellow stars, and red dashed line denote position of the MIT estimated by four different methods. Blue triangles denote when the ground state charge gap falls below a threshold of 0.1. Bright green triangles denote critical $t_W^*$ inferred by performing a linear fit to $\Delta(N_e, U) = a U - b$ and using the $x$-intercept of the fit as $U_c/t_W$~\cite{Morales-Duran2021,Botzung2024a} (for example fits, see \cref{fig:Uc-fit}). 
    Yellow stars denote when the charge gap averaged over twisted boundary conditions reaches 0~\cite{Koretsune2007} (see \cref{fig:gap-vs-tW-KM}). The red dashed line is an analytical estimate of the MIT based on \cref{eq:Weff}. \textbf{c} Ground state spin configuration. \textbf{d} Magnitude of band-projected spin correlation function $C^x(\vec{K})$ in the $S_z=0$ sector. Green dashed line in \textbf{d} (identical to that in \textbf{a}) is a guide to the eye. Identical magenta dashed lines in \textbf{a}, \textbf{c} and \textbf{d} denotes where the nearest neighbor exchange $J^{zz}_1(M,t_W)$ changes sign as per \cref{eq:J-total}. ED results are obtained on a hexagonal \texttt{12d4} Betts cluster~\cite{Betts1996,Betts1999,Gustiani2015} with $t'=0.05$, $M > 3\sqrt{3}t' = 0.26$. }
    \label{fig:km-phase}
\end{figure*}

The ED ground state phase diagram for the quarter-filled interacting Kane-Mele model is shown in \cref{fig:km-phase}\textbf{a}. Depending on $t_W$ and $M$, the model exhibits Mott FM, Mott AFM, and metallic ground states, as well as an insulating region with ambiguous spin order signatures.
The phase diagram is drawn based on combined analysis of the ground state charge gap $\Delta(N_e) = \Delta(N_e +1) + \Delta(N_e-1) - 2\Delta(N_e)$ (shown in \cref{fig:km-phase}\textbf{b}), spin configuration (shown in \cref{fig:km-phase}\textbf{c}), and band-projected spin correlation functions, summed over both orbitals (defined in \cref{sub:band_projected_spin_correlation_functions} and shown in \cref{fig:km-phase}\textbf{d}). $N_e$ denotes the number of electrons in the system, which at quarter-filling, is equal to the number of unit cells in the cluster.

We first analyze charge order. As we expect, at each fixed $M$, the system is insulating at small $t_W$ and metallic at larger $t_W$. Because it is challenging to pinpoint the exact location of a metal-to-insulator transition in finite-size simulations, we use four criteria to numerically locate the position of the Mott MIT in \cref{fig:km-phase}\textbf{b}.  While each method produces slightly different estimates, the overall trend is clear: at fixed $t_W$, decreasing $M$ will drive the system from a Mott insulator to a metal, indicating that larger quantum geometry obstructs the localization of charge and favors metallicity, as we expect. Previous ED and DMRG studies of the single-band triangular lattice Hubbard model (equivalent to setting $M\rightarrow \infty$ in the Kane-Mele model) shows that the model undergoes a MIT at $t_W/U \sim 1/8.5 \approx 0.12$~\cite{Szasz2020,Botzung2024a}, possibly becoming a spin liquid before becoming a conventionally ordered $120^\circ$ AFM at $t_W/U \sim 1/10.6 \approx 0.09$. 
The approximate MIT we observe has positive slope and approach this trivial, single-band limit as $M$ is increased. 

Next, we analyze spin order. Due to the $\mathbb{Z}_2 \times U(1)$ spin symmetry of the model, we must consider both in-plane and out-of-plane spin order. \cref{fig:km-phase}\textbf{c} shows that at small $t_W$, within the Mott insulating phase, the spin order is out-of-plane $\mathbb{Z}_2$ FM, which we identify as the Mott FM~\cite{Zhang2019,Tasaki2020}. To further disambiguate spin ordering tendencies within the $S_z=0$ sector, we calculate the out-of-plane and in-plane band-projected spin correlations $C^z(\vec{q})$ and $C^x(\vec{q})$, as defined in \cref{sub:band_projected_spin_correlation_functions}. We find that the spin correlations to be almost isotropic (\cref{fig:km-spincorr-afm,fig:km-spincorr-metal,fig:km-high-sym-12d4}), but with slightly stronger in-plane correlations at $\vec{q} = \vec{K}$, shown in \cref{fig:km-phase}\textbf{d}. The region of strongly peaked $C^x(\vec{K})$ falls within the Mott insulating phase, so we identify it as the Mott AFM, with conventional $120^\circ$ in-plane Ne\'el order. Finally, there is a remnant region within the Mott insulating phase with weaker spin correlations that resists easy identification in finite-size simulations, shown in yellow in \cref{fig:km-phase}\textbf{a}. Crossing the boundary of this region via increasing $t_W$ within the Mott phase coincides with a discontinuous drop in spin correlations shown in \cref{fig:km-phase}\textbf{d}, which subsequently persist deep into the metallic phase for large $t_W$. While inaccessible in our exact diagonalization simulations, we note that prior DMRG calculations of the single-band triangular lattice Hubbard model \cite{Koretsune2007,Szasz2020,Wietek2021} have found evidence for a chiral spin liquid emerging as a true intervening phase between metal and Mott AFM in the thermodynamic limit. 

The observed phase boundary between Mott FM and Mott AFM phases corroborates recent proposals on quantum-geometry-driven FM-AFM transitions in narrow bands~\cite{Hu2025a,Repellin2020a}. Increasing $t_W$ at fixed $M$ induces a FM-AFM transition; at fixed nonzero but small $t_W$, increasing $M$ also induces a FM-AFM transition. 
While a Mott FM-AFM transition was also previously observed in Ref.~\cite{Hu2021,Morales-Duran2022a}, is was tuned by continuum model parameters, without explicitly separating quantum geometrical and bandwidth effects.

Both the quantum-geometric Mott transition and FM-AFM transition may be intuitively understood by considering the Wannier functions of the partially filled band. The quantum geometry of Bloch states controls the minimal spread of Wannier orbitals, which affects the two-particle projected interaction elements written in a basis with non-atomic Wannier orbitals~\cite{Marzari2012,Morales-Duran2022a}. For the Mott transition, delocalization of the Wannier orbital lowers the the effective on-site interaction and introduces density-assisted hopping and nearest-neighbor repulsion, so that a larger bare Hubbard interaction is required to form a Mott insulator. For the FM-AFM transition, delocalization of the Wannier orbital
increases the nearest-neighbor direct exchange, which tends to be ferromagnetic. This direct exchange competes with the nearest-neighbor antiferromagnetic superexchange, which is controlled by the bare band dispersion and density-assisted hopping.

To make the above arguments more quantitative, we rewrite the interaction term in \cref{eq:proj-final-momentum} as 
\begin{equation}
    \hat{V}_{\mathrm{eff}} =
    \sum_{ijkl}
    U_{ijkl}
    b_{i\uparrow}^\dagger 
    b_{j\downarrow}^\dagger 
    b_{k\downarrow} 
    b_{l\uparrow} \label{eq:V-wannier}
\end{equation}
where $b^\dagger_{is}$ creates an electron of spin $s$ in the Wannier orbital centered at site $i$, and the prefactors in \cref{eq:V-wannier} are two-particle matrix elements 
\begin{multline}
    U_{ijkl}=\langle ij | \hat{V}_\mathrm{eff}|  kl \rangle \\
    = \left[\sum_{\alpha} \int d\vec{r} W^*_{i\uparrow\alpha}(\vec{r}) W^*_{j\downarrow\alpha}(\vec{r}) W_{k\downarrow\alpha}(\vec{r}) W_{l\uparrow\alpha}(\vec{r})\right] \label{eq:U-from-W}
\end{multline}
where $W_{is\alpha}(\vec{r})$ denotes the real space wavefunction for the orbital $\alpha$ component of the localized Wannier function centered at site $i$ for an electron with spin $s$.  As the two-band model $\hat{H}_0$ we start with is exactly solvable, we can explicitly write down $\ket{u_n^s(\vec{k})}$ which are smooth in the entire Brillouin Zone, which allows us to simply take a Fourier transform (without extra phase factors), and obtain smooth and exponentially localized Wannier functions. We adapt the gauge choice
\begin{equation}
\ket{u^\uparrow_{n=0}(\vec{k})} =
\frac{1}{\sqrt{2|\vec{a}| (|\vec{a}|-a_3)}} \begin{bmatrix}
a_3 - |\vec{a}|\\
a_1 + ia_2
\end{bmatrix},
\label{eq:2b-uB}
\end{equation}
which produces almost maximally-localized Wannier functions, as shown in \cref{fig:gauge-choice-KM}. A typical Wannier function for the Kane-Mele model using this gauge choice is shown in \cref{fig:km-Wannier}\textbf{a}.

\begin{figure}[t]
    \centering
    \includegraphics[width=0.45\textwidth]{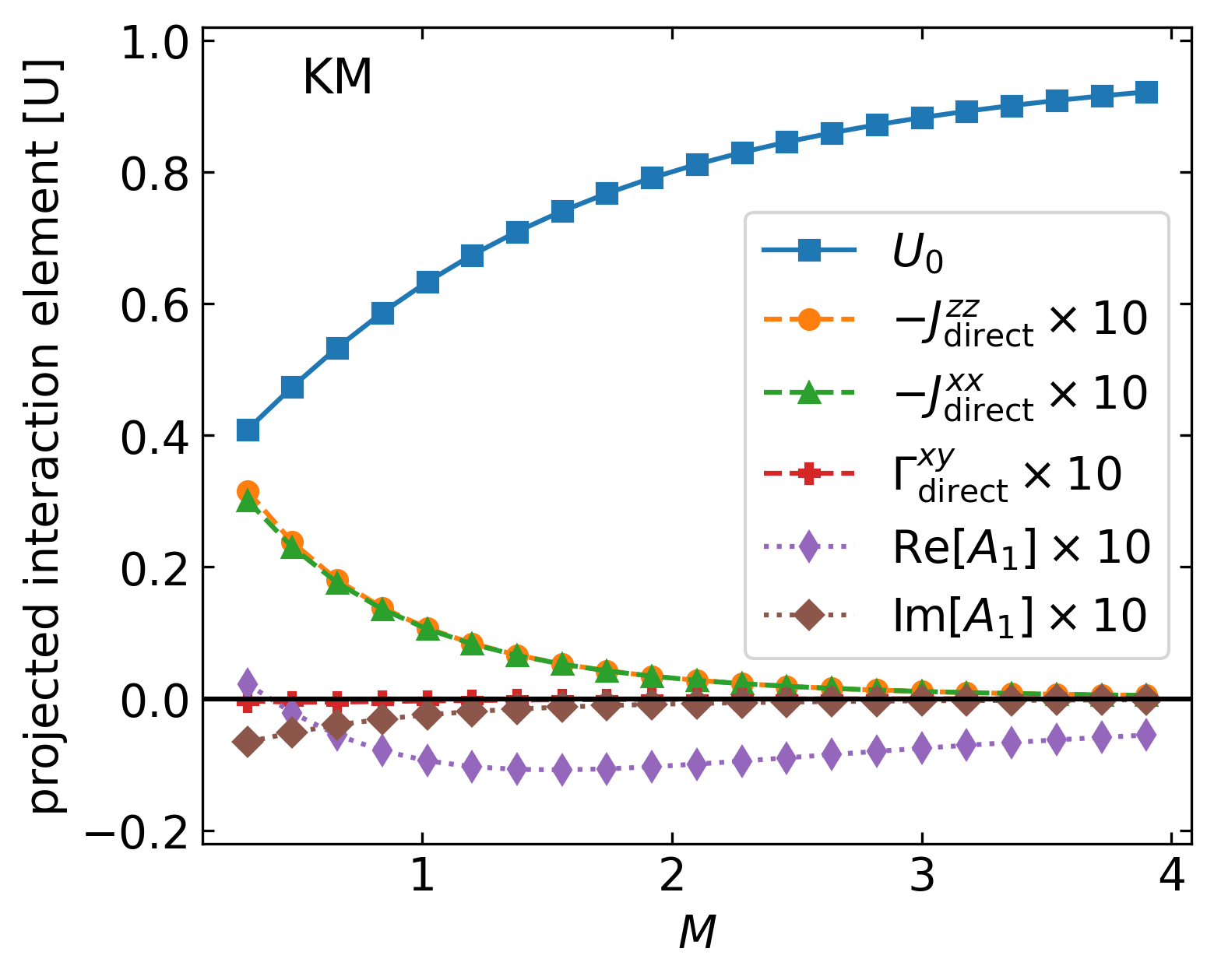}
    \caption{Nearest-neighbor interaction matrix elements for Wannier orbitals displaced by $\vec{a}_1$ for the Kane-Mele model, expressed in terms of bare interaction strength $U$.
    Except $U_0$, all other interaction terms have been re-scaled to improve visibility. Effective nearest-neighbor repulsion and pair hopping are related to spin exchange matrix elements via $U_1 = - J^{zz}_{\mathrm{direct}}/4$, and $P_1 = -J^{zz}_{\mathrm{direct}}/2$.}
    \label{fig:int-short-KM}
\end{figure}

Assuming that Wannier functions decay fast enough to justify keeping only local and nearest neighbor terms in \cref{eq:V-wannier}, we find the interaction term can be rewritten as an extended Hubbard Hamiltonian
\begin{align}
    \hat{V}_{\mathrm{eff}} &\approx
    U_0\sum_i  n_{i\uparrow} n_{i\downarrow} 
    + 
    U_1\sum_{\expval{ij}} n_i n_j \nonumber \\
    &+ 
    \sum_{\expval{ij}} \left[J^{zz}_{\mathrm{direct}} S^z_i S^z_j + J^{xx}_{\mathrm{direct}} (S^x_i S^x_j + S^y_i S^y_j) \right. \nonumber \\ 
    & \quad \left.+ \Gamma^{xy}_{\mathrm{direct}} (S^x_i S^y_j - S^y_i S^x_j)\right] \nonumber\\
    &+ 
    \sum_{\langle ij \rangle} 
    \left[A_1 (n_{i\downarrow} + n_{j\downarrow})
    b_{i\uparrow}^\dagger 
    b_{j\uparrow}
    +
    \mathrm{h.c.}\right]\nonumber\\
    &+ 
    \sum_{\langle ij \rangle} 
    \left[
    A_1^* 
    (n_{i\uparrow} + n_{j\uparrow})
    b_{i\downarrow}^\dagger 
    b_{j\downarrow} +
    \mathrm{h.c.}\right]\nonumber\\
    &+ 
    \sum_{\langle ij \rangle} 
    P_1\left[
    \Delta_i^\dagger
    \Delta_j + \mathrm{h.c.}\right] \label{eq:short-range-H}
\end{align}
where the local density $n_{i(s)}$, spin $S_i^{\gamma}$, and pair creation/annihilation $\Delta_i^{(\dagger)}$ operators are defined in terms of Wannier orbital creation and annhilation operators (see \cref{sub:local_operators_in_wannier_basis}). The local Hubbard and nearest-neighbor density interaction strengths are
\begin{equation}
    U_0 = U_{iiii}, \  U_1 = \frac{1}{2}U_{ijji}
\end{equation}
direct spin exchange interactions are
\begin{multline}
    J^{zz}_{\mathrm{direct}} = -2U_{ijji}, \ 
    J^{xx}_{\mathrm{direct}} = -U_{ijij} - U_{jiji}, \\
    \Gamma^{xy}_{\mathrm{direct}} = i \left(U_{ijij} - U_{jiji}\right)
\end{multline}
and density-assisted hopping and pair hopping coefficients are
\begin{equation}
    A_1 = U_{iiij}, \ P_1 = U_{iijj} = U_{ijji}.
\end{equation}
Dependence of these short-range interaction elements on orbital imbalance $M$ is shown in \cref{fig:int-short-KM}.

Using \cref{eq:U-from-W,eq:short-range-H}, we can make an analytical estimate of the location of the Mott transition based on calculated values of $U_0, U_1$ and $A_1$ in the Wannier basis. The best pure Hubbard approximation to system with small nearest neighbor repulsion has renormalized on-site interaction strength~\cite{Schuler2013a} $U^* = U_0-U_1$, 
and the total bandwidth after including effects of density-assisted hopping is approximately $W(t_W,A_1) \approx 9(t_W + \abs{A_1})$.
Using results from the triangular lattice Hubbard model~\cite{Szasz2020}, we estimate the MIT occurs at 
\begin{equation}
    U^*/W(t_W,A_1) = 8.5/9 \label{eq:Weff}
\end{equation}
This estimate is drawn as a red dashed line in \cref{fig:km-phase}\textbf{b}, and agrees fairly well with numerical estimates of the MIT. 

Using \cref{eq:U-from-W,eq:short-range-H}, we can also make an analytical estimate of the location of the FM-AFM transition. Assuming that $U_0$ is much greater than $U_1, P_1, A_1, t_W$ and direct exchange terms, the low-energy Hilbert space consists of singly occupied sites, and the Hamiltonian that governs remaining spin degrees of freedom may be found by performing a Schrieffer-Wolff transform. 
To second order in $(t_W+A_1)/U_0$, the effective short range spin exchange Hamiltonian may be written as a nearest-neighbor XXZ model
\begin{multline}
    \hat{H}_{\mathrm{spin}} = \sum_{\expval{ij}} \left[J^{zz}_{1} S^z_i S^z_j + J^{xx}_{1} (S^x_i S^x_j + S^y_i S^y_j) \right.\\
    \left.+ \Gamma^{xy}_{1} (S^x_i S^y_j - S^y_i S^x_j)\right]
\end{multline}
where
\begin{multline}
    J_1^{zz} = J^{zz}_{\mathrm{direct}} + J_{\mathrm{super}}^{zz}, \quad J_1^{xx} = J^{xx}_{\mathrm{direct}} + J_{\mathrm{super}}^{xx}, \\
    \Gamma_1^{xy} = \Gamma^{xy}_{\mathrm{direct}} + \Gamma_{\mathrm{super}}, \label{eq:J-total}
\end{multline}
and the super-exchange terms include contributions from both non-interacting hopping $t_W$ and density-assisted hopping $A_1 = |A|\exp[i\phi]$:
\begin{align}
    J_{\mathrm{super}}^{zz} &= 
    \frac{4 (t_W^2 - 2\abs{A} t_W \cos\phi + \abs{A}^2)}{U_0 - U_1}, \\
    J_{\mathrm{super}}^{xx} &= \frac{4 (t_W^2 - 2\abs{A} t_W \cos\phi + \abs{A}^2 \cos(2\phi))}{U_0 - U_1}\label{eq:J-super}.
\end{align}

Empirically, we find the diagonal exchange terms $J^{zz}_{1}$ and $J^{xx}_{1}$ to be almost isotropic, and the symmetric off-diagonal exchange $\Gamma_{1}^{xy}$ to be much smaller than both diagonal exchange terms. Assuming that it is safe to ignore longer range exchange terms (as Wannier functions decay exponentially, this is a reasonable assumption), we use the sign of the nearest neighbor exchange $J_1^{zz}(M,t_W)$
to estimate the location of the FM-AFM transition. 
This estimate is shown as magenta dashed lines in \cref{fig:km-phase}\textbf{c} and \cref{fig:km-phase}\textbf{d}, and agrees well with the numerically determined phase boundary. 

Additionally, we note that \cref{fig:km-phase}\textbf{a} suggests that when quantum geometric effects are large, it is possible to observe a direct Mott FM to metal transition, without crossing an intervening Mott AFM phase, contrary to expectations based on the single-band Hubbard model. 

Finally, to ensure that the quantum-geometric Mott metal-to-insulator and FM-AFM transitions are not isolated to a single model Hamiltonian, and to make closer contact with the standard square lattice Hubbard model, we also performed simulations on the BHZ-Hubbard model~\cite{Qi2006,Tzeng2023,Mai2023c}. Results on the BHZ-Hubbard model are shown in \cref{sec:additional_results_in_the_bhz_hubbard_model}, and qualitatively agree with our findings in the Kane-Mele-Hubbard model.

\section{Conclusion and Outlook} 
\label{sec:conclusions}
In summary, we demonstrate that quantum geometry can drive a Mott metal-to-insulator transition and tune the Mott insulator between ferromagnetic and antiferromagnetic phases. Augmenting conventional Hubbard models with quantum geometry prevents complete charge localization and enriches the resulting phase diagram, realizing a Bloch band geometry-driven mechanism for Mott metal-to-insulator transitions that is distinct from bandwidth-driven and doping-driven MITs. 
We detail the mechanism for quantum geometry-driven phase transitions by tracking the evolution of dominant Coulomb scattering matrix elements as a function of the real-space spread of a basis of exponentially- but poorly-localized Wannier functions, and find good agreement between real-space estimates and numerical results in the Kane-Mele-Hubbard model with a half-filled valence band.

Our results establish quantum geometry as a powerful independent ingredient in topologically-trivial correlated electron systems, which is distinct from kinetic considerations and can enrich Mott metal-insulator transitions in multi-orbital and moir\'e systems. This suggests that quantitative analyses of Mott physics and MITs should account for the Bloch wavefunction structure of the partially filled valence bands beyond effective local Hubbard models. Our work provides a foundation for a more analytical and systematic understanding of how quantum geometry and Wannier localization affects phases in strongly correlated materials, and points to the broader existence of other quantum-geometry-driven phase transitions of correlated electrons beyond topological or fractionalized states of matter.

Several directions follow naturally. The presence of a remnant region in the Mott insulator in our Kane-Mele model results suggest that quantum geometric effects may enhance the stability of the chiral spin liquid phase in the triangular lattice Hubbard model. Imperfectly localized Wannier orbitals may induce longer range (beyond nearest-neighbor) spin exchange interactions, which are believed to frustrate AFM ordering and favor the formation of a chiral spin liquid~\cite{Cookmeyer2021}. 
It would also be interesting to identify spectral and transport signatures that distinguish quantum-geometry-driven Mott transitions from conventional bandwidth-driven ones~\cite{Brinkman1970,Imada1998,Senthil2008,Kohno2012,Walsh2019}. For this task, it willbe fruitful to recontextualize known results in the extended Hubbard model in terms of quantum geometric effects, as density-assisted hopping~\cite{Kovalska2025} and pair hopping terms~\cite{Robaszkiewicz1999} are known to modify the Mott MIT. Improving our understanding of mechanisms controlling the Mott transition and proximate quantum phases will also benefit device applications that exploit ``Mott switching'' for ultrafast control of conductivity and magnetism~\cite{Milloch2024}.

\section{Data and Code Availability}

Aggregated numerical data and analysis routines required to reproduce the figures can be found at \url{10.5281/zenodo.17117151}. Raw simulation data that support the findings of this study are stored on the M\"obius cluster at the University of Pennsylvania and are available from the corresponding author upon reasonable request.

\section*{Acknowledgements}
JKD is indebted to Terrence Tai and Brandon Monsen for help with ED code. We thank Stefan Divic and Sandeep Joy for feedback on the manuscript and discussions. We also thank Nishchal Verma, Patrick Ledwith, and the UPenn CMT group at large for helpful discussions.

This work was supported by the U.S. Department of Energy, Office of Basic Energy Sciences, Early Career Award Program, under Award No. DE-SC0024494.
Computational work was performed on the M\"obius cluster at the University of Pennsylvania and on resources of the National Energy Research Scientific Computing Center (NERSC), a Department of Energy User Facility using NERSC award BES-ERCAP0032059.
\bibliography{main}

\clearpage

\appendix
\onecolumngrid
\renewcommand{\thefigure}{S\arabic{figure}}
\renewcommand{\thetable}{S\arabic{table}}
\setcounter{figure}{0} 
\setcounter{table}{0}  

\section{Additional definitions}
\label{sec:defns}

\subsection{Band-projected spin correlation functions}
\label{sub:band_projected_spin_correlation_functions}

Orbitally-resolved local spin operators are gauge invariant observables expressed using fermion creation and annihilation operators in real space $c_{i\alpha \mu}^\dagger, c_{i\alpha \nu}$ and Pauli matrices $\sigma^\gamma_{\mu\nu}$, where $\gamma \in \{x,y,z\}$, $i$ indexes sites, $\alpha \in \{A,B\}$ indexes orbitals,  $\mu,\nu \in \{\uparrow,\downarrow\}$ indexes spin. The spin operator for orbital $\alpha$ in unit cell $i$ is
\begin{equation}
    S^\gamma_{i\alpha} = \frac{\hbar}{2} 
    \begin{pmatrix}
        c_{i\alpha\uparrow}^\dagger &
        c_{i\alpha\downarrow}^\dagger
        \end{pmatrix}\sigma^\gamma \begin{pmatrix}
        c_{i\alpha\uparrow}\\
        c_{i\alpha\downarrow} 
    \end{pmatrix} 
    = \frac{\hbar}{2} \sum_{\mu\nu} c_{i\alpha\mu}^\dagger \sigma^\gamma_{\mu\nu} c_{i\alpha\nu}\quad
    \begin{cases}
    S_{i\alpha}^x= \frac{\hbar}{2}\left(c^\dagger_{i\alpha\uparrow}c_{i\alpha\downarrow} + c^\dagger_{i\alpha\downarrow}c_{i\alpha\uparrow}\right) \\
    S_{i\alpha}^y= -\frac{i\hbar}{2}\left(c^\dagger_{i\alpha\uparrow}c_{i\alpha\downarrow} - c^\dagger_{i\alpha\downarrow}c_{i\alpha\uparrow}\right) \\
    S_{i\alpha}^z= \frac{\hbar}{2}\left(c^\dagger_{i\alpha\uparrow}c_{i\alpha\uparrow} - c^\dagger_{i\alpha\downarrow}c_{i\alpha\downarrow}\right) 
    \end{cases} \label{eq:orbital-spin}
\end{equation}
Adapting the following convention for Fourier transforms of bosonic operators,
\begin{align}
    S_\alpha^z(\vec{q}) = \frac{1}{\sqrt{N}}\sum_{\vec{R}_i} \exp[-i \vec{q} \cdot (\vec{R}_i + \bm{\tau}_\alpha)] S_{i\alpha}^z \nonumber\\
    S_\alpha^z(-\vec{q}) = \frac{1}{\sqrt{N}}\sum_{\vec{R}_j} \exp[i \vec{q} \cdot (\vec{R}_j + \bm{\tau}_\alpha)] S_{j\alpha}^z \label{eq:spin-rk},
\end{align}
the (orbital $\alpha$, momentum $\vec{q}$) resolved spin operators are written as
\begin{equation}
    S^\gamma_{\alpha}(\vec{q}) 
    = \frac{\hbar}{2} \sum_{\vec{k}} \sum_{\mu\nu} c_{\vec{k}+\vec{q}\alpha\mu}^\dagger \sigma^\gamma_{\mu\nu} c_{\vec{k}\alpha\nu} \quad \begin{cases}
    S_\alpha^x(\vec{q}) 
    = \frac{\hbar}{2}\sum_{\vec{k}} 
    \left(c_{\vec{k}+\vec{q}\alpha\uparrow}^\dagger c_{\vec{k}\alpha\downarrow} + c_{\vec{k}+\vec{q}\alpha\downarrow}^\dagger c_{\vec{k}\alpha\uparrow}\right) \\
    S_\alpha^y(\vec{q}) 
    = -\frac{i\hbar}{2}\sum_{\vec{k}} 
    \left(c_{\vec{k}+\vec{q}\alpha\uparrow}^\dagger 
    c_{\vec{k}\alpha\downarrow} - 
    c_{\vec{k}+\vec{q}\alpha\downarrow}^\dagger 
    c_{\vec{k}\alpha\uparrow}\right) \\ 
    S_\alpha^z(\vec{q}) 
    = \frac{\hbar}{2}\sum_{\vec{k}} 
    \left(c_{\vec{k}+\vec{q}\alpha\uparrow}^\dagger c_{\vec{k}\alpha\uparrow} - c_{\alpha\vec{k}+\vec{q}\downarrow}^\dagger c_{\vec{k}\alpha\downarrow}\right)
    \end{cases}. \label{eq:Sagq}
\end{equation}

Using the following relationship between band creation/annihilation operators and orbital creation/annihilation operators [c.f. \cref{eq:uank}]
\begin{align}
    b_{n\vec{k}} &= \sum_{\alpha}c_{\vec{k}\alpha} u^*_{\alpha n}(\vec{k}) \quad
    & c_{\vec{k}\alpha} = \sum_{n} u_{\alpha n}(\vec{k}) b_{n\vec{k}} \label{eq:bk-ck}\\
    b_{n\vec{k}}^\dagger &= \sum_{\alpha}c_{\vec{k}\alpha}^\dagger u_{\alpha n}(\vec{k}) \quad
    & c_{\vec{k}\alpha}^\dagger = \sum_{n}u^*_{\alpha n}(\vec{k}) b_{n\vec{k}}^\dagger \label{eq:bk-ck-d}
\end{align}
We may rewrite \cref{eq:Sagq} in terms of band creation and annihilation operators:
\begin{equation}
    S^\gamma_{\alpha}(\vec{q}) 
    = \frac{\hbar}{2} \sum_{\vec{k}} \sum_{\mu\nu} \sum_{nn'} u^*_{\alpha n \mu }(\vec{k}+\vec{q}) b_{n\vec{k}+\vec{q}\mu}^\dagger  \sigma^\gamma_{\mu\nu} u_{\alpha n' \nu}(\vec{k}) b_{n'\vec{k}\nu}.
\end{equation}

Thus, the orbitally-resolved spin-spin correlation function is expressed as
\begin{equation}
    S^\gamma_{\alpha}(\vec{q}) S^{\gamma'}_{\beta}(-\vec{q}) 
    = \frac{\hbar^2}{4} 
    \sum_{\vec{k}\vec{k'}} 
    \sum_{\substack{s_1s_2\\s_3s_4}} 
    \sum_{\substack{n_1 n_2\\n_3 n_4 }} 
    u^*_{\alpha n_1 s_1 }(\vec{k}+\vec{q}) 
    u_{\alpha n_2 s_2}(\vec{k}) 
    u^*_{\beta n_3 s_3 }(\vec{k}'-\vec{q})
    u_{\beta n_4  s_4}(\vec{k}') 
    \sigma^\gamma_{s_1s_2} 
    \sigma^{\gamma'}_{s_3s_4}
    b_{n_1 \vec{k}+\vec{q}s_1}^\dagger     
    b_{n_2 \vec{k}s_2}  
    b_{n_3 \vec{k}'-\vec{q}s_3}^\dagger  
    b_{n_4 \vec{k}'s_4}
\end{equation}

First normal ordering all band creation and annihilation operators, then band projecting into a single band $n_1=n_2=n_3=n_4=0$ (and omitting band indices), we find the orbitally-resolved band-projected spin-spin correlation to be expressed as
\begin{align}
    C_{\alpha\beta}^{\gamma\gamma'}(\vec{q}) &= \mathcal{P} \colon S^\gamma_{\alpha}(\vec{q})S^{\gamma'}_{\beta}(\vec{-q}) \colon \mathcal{P} \\
    &= \frac{\hbar^2}{4} 
    \sum_{\vec{k}\vec{k'}} 
    \sum_{\substack{s_1s_2\\s_3s_4}} 
    u^*_{\alpha  s_1 }(\vec{k}+\vec{q}) 
    u_{\alpha  s_2}(\vec{k}) 
    u^*_{\beta  s_3 }(\vec{k}'-\vec{q})
    u_{\beta   s_4}(\vec{k}') 
    b_{ \vec{k}+\vec{q}s_1}^\dagger    
    b_{ \vec{k}'-\vec{q}s_3}^\dagger  
    b_{ \vec{k}'s_4} 
    b_{ \vec{k}s_2}  \\
    & \quad + \frac{\hbar^2}{4}
    \delta_{\alpha\beta}
    \sum_{\vec{k}} 
    \sum_{s_1s_2s_4}
    \sigma^\gamma_{s_1s_2} 
    \sigma^{\gamma'}_{s_2s_4}
    u^*_{\alpha  s_1 }(\vec{k} ) 
    u_{\alpha   s_4}(\vec{k} ) 
    b_{ \vec{k} s_1}^\dagger     
    b_{ \vec{k} s_4}
\end{align}
To detect in-plane and out-of-plane spin order, we use $\gamma=\gamma'=z$ and $\gamma=\gamma'=x$ respectively. Additionally, because we focus on overall quarter filling, we are not interested in spin correlations within one unit cell, so we only calculate orbitally diagonal correlations with $\alpha = \beta$.
The spin correlations shown in the main text are the sums of orbitally diagonal correlations
\begin{equation}
    C^{\gamma}(\vec{q}) \equiv C^{\gamma\gamma}_{AA}(\vec{q}) + C^{\gamma\gamma}_{BB}(\vec{q})
\end{equation}
with $\gamma \in {x,z}$.

\subsection{Local operators in Wannier basis}
\label{sub:local_operators_in_wannier_basis}
Explicitly, local density, spin and pair operators in \cref{eq:short-range-H} are defined using Wannier orbital creation and annihilation operators as follows:
\begin{equation}
    n_{js} =  b^\dagger_{js}b_{js}, 
    \quad n_j = \sum_s n_{js}, 
    \quad S^\gamma_{j} = \frac{\hbar}{2}\sum_{ss'} b_{js}^\dagger \sigma^{\gamma}_{ss'} b_{js'},  
    \quad \Delta_j = b_{j\downarrow}b_{j\uparrow} \label{eq:nSD-defn}
\end{equation}
where $b^\dagger_{is}$ creates an electron of spin $s$ in the Wannier orbital centered at site $i$, $s\in \{\downarrow,\uparrow\}$, $\alpha\in\{x,y,z\}$, and $\sigma$ denote Pauli matrices. We have omitted the band index $n=0$. 

Here, we emphasize that despite similar notation, the local spin operator in Wannier basis defined in \cref{eq:nSD-defn} should not be confused with the orbitally-resolved local spin operator defined in \cref{eq:orbital-spin}. The latter is defined in terms of the bare orbitals of the two band model and is thus gauge-independent, while the former relies on Wannier function choice and is thus gauge-dependent.

\section{Supplementary Plots for the Kane-Mele-Hubbard Model}
\label{sec:supplementary_plots}

\begin{figure}[htbp]
    \centering
    \includegraphics[width=0.5\textwidth]{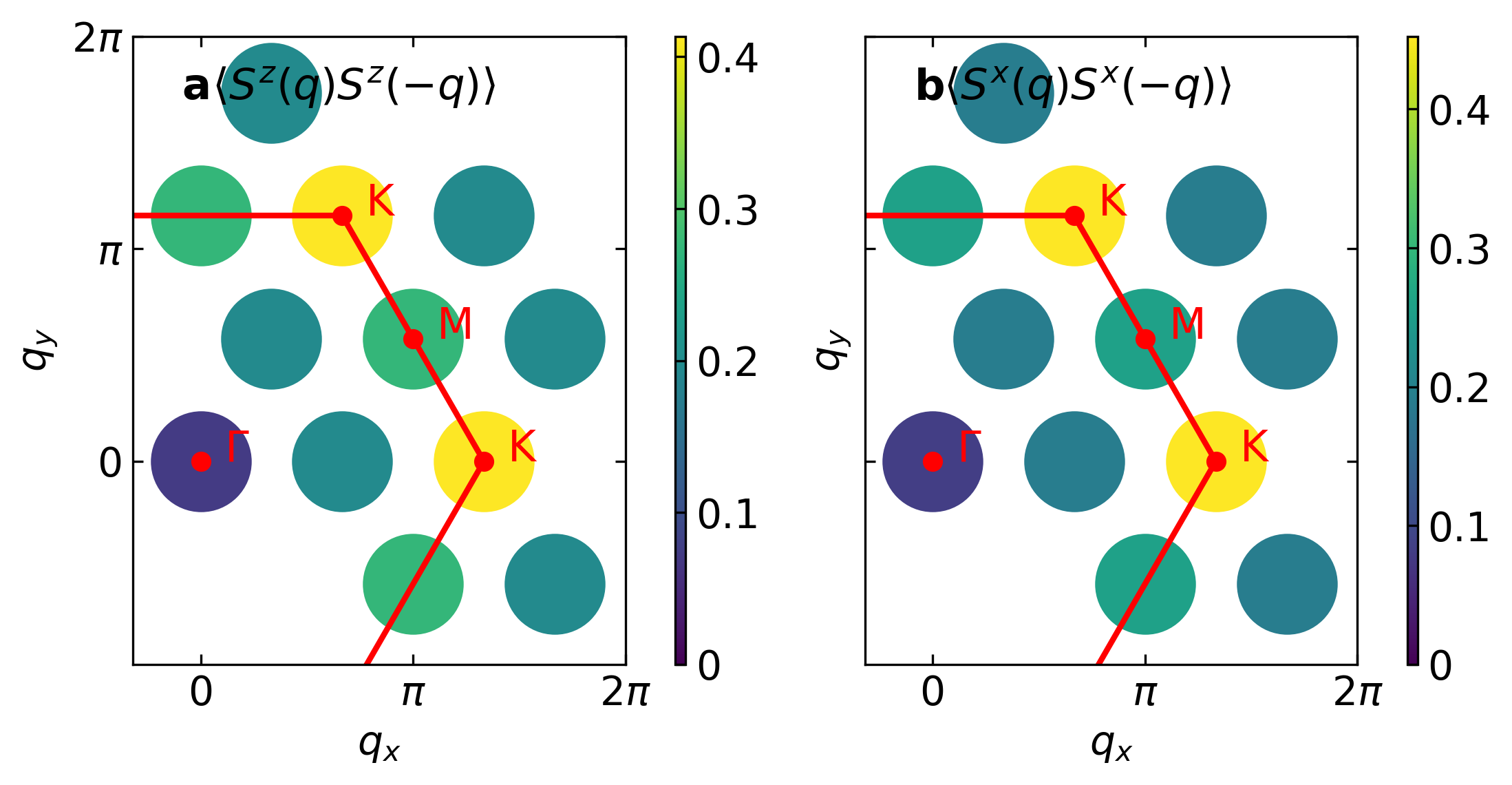}
    \caption{Band-projected spin correlation functions of Kane-Mele model in Mott AFM phase on \texttt{12d4} cluster, demonstrating $120^\circ$ in-plane Ne\'el order. Parameters: $t'=0.05$, $M=1.4$, $t_W = 0.04$, $U=1$.}
    \label{fig:km-spincorr-afm}
\end{figure}

\begin{figure}[htbp]
    \centering
    \includegraphics[width=0.5\textwidth]{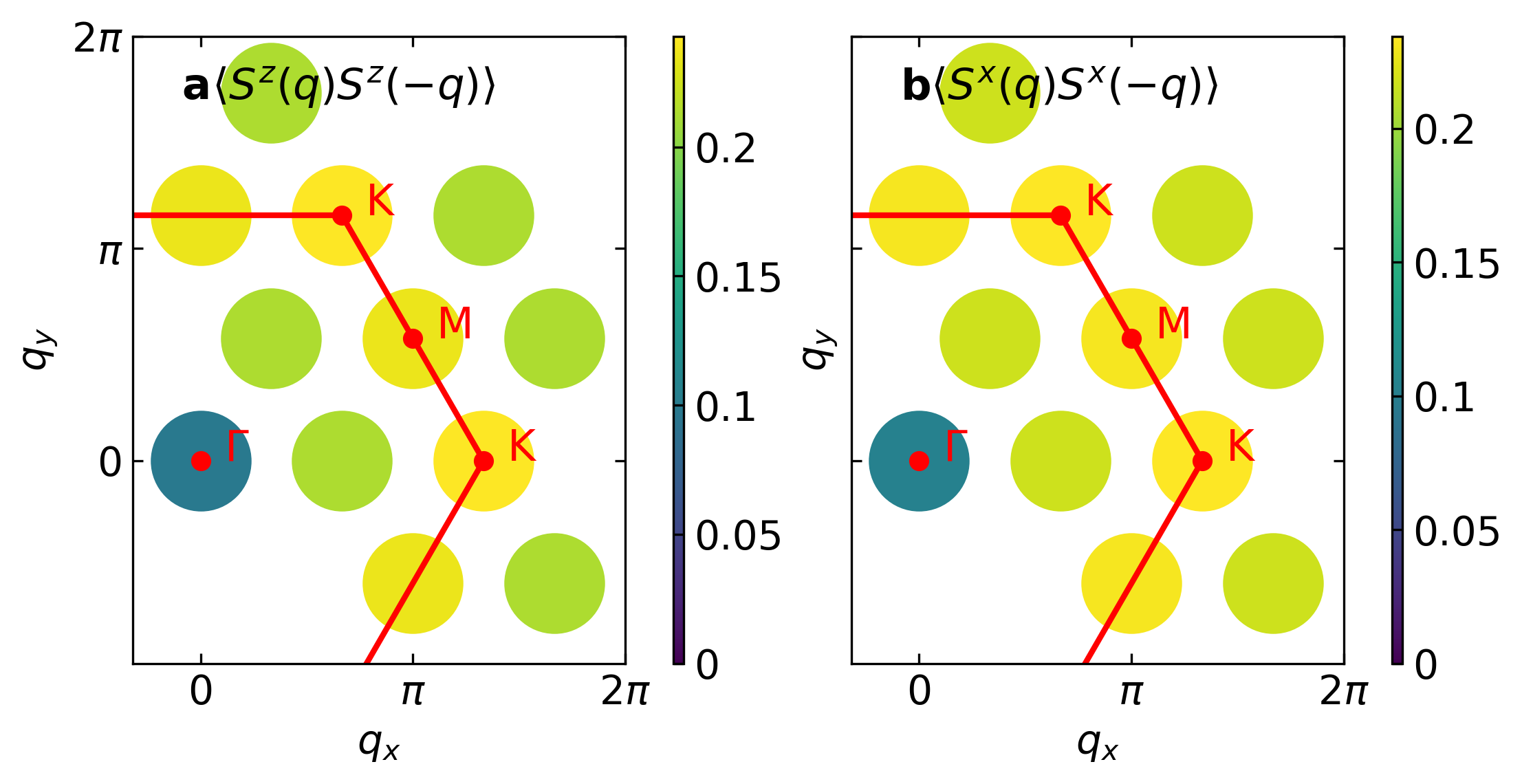}
    \caption{Band-projected spin correlation functions of Kane-Mele model in metallic phase on \texttt{12d4} cluster, demonstrating paramagnetic spin correlations. Parameters: $t'=0.05$, $M=1$, $t_W = 0.1$, $U=1$.}
    \label{fig:km-spincorr-metal}
\end{figure}

\begin{figure}[htbp]
    \centering{}
    \includegraphics[width=0.7\textwidth]{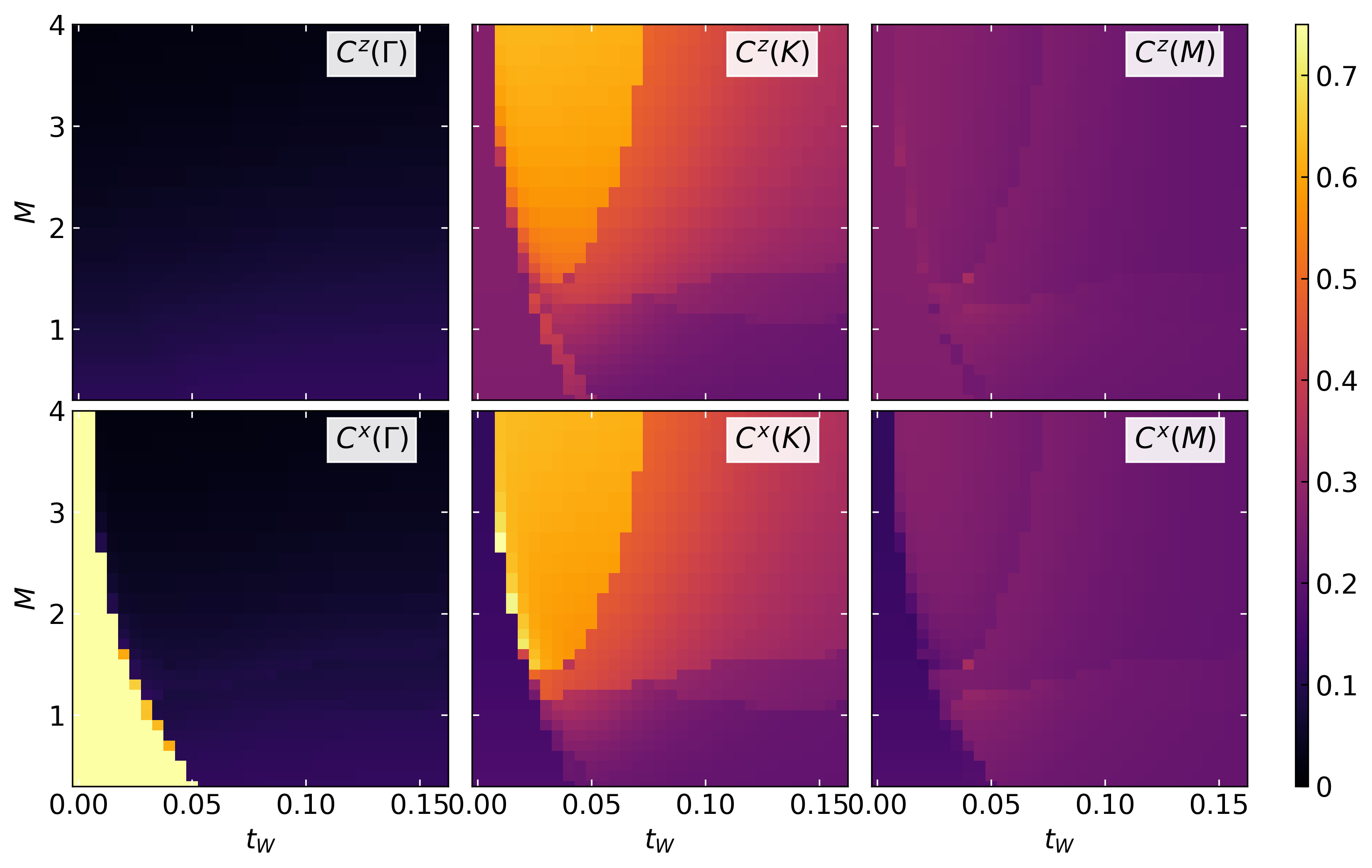}
    \caption{Projected spin correlation functions of the Kane-Mele model in the $S_z = 0$ sector at high symmetry points of the \texttt{12d4} cluster.}
    \label{fig:km-high-sym-12d4}
\end{figure}

\begin{figure}[htbp]
    \centering
    \includegraphics[width=0.55\textwidth]{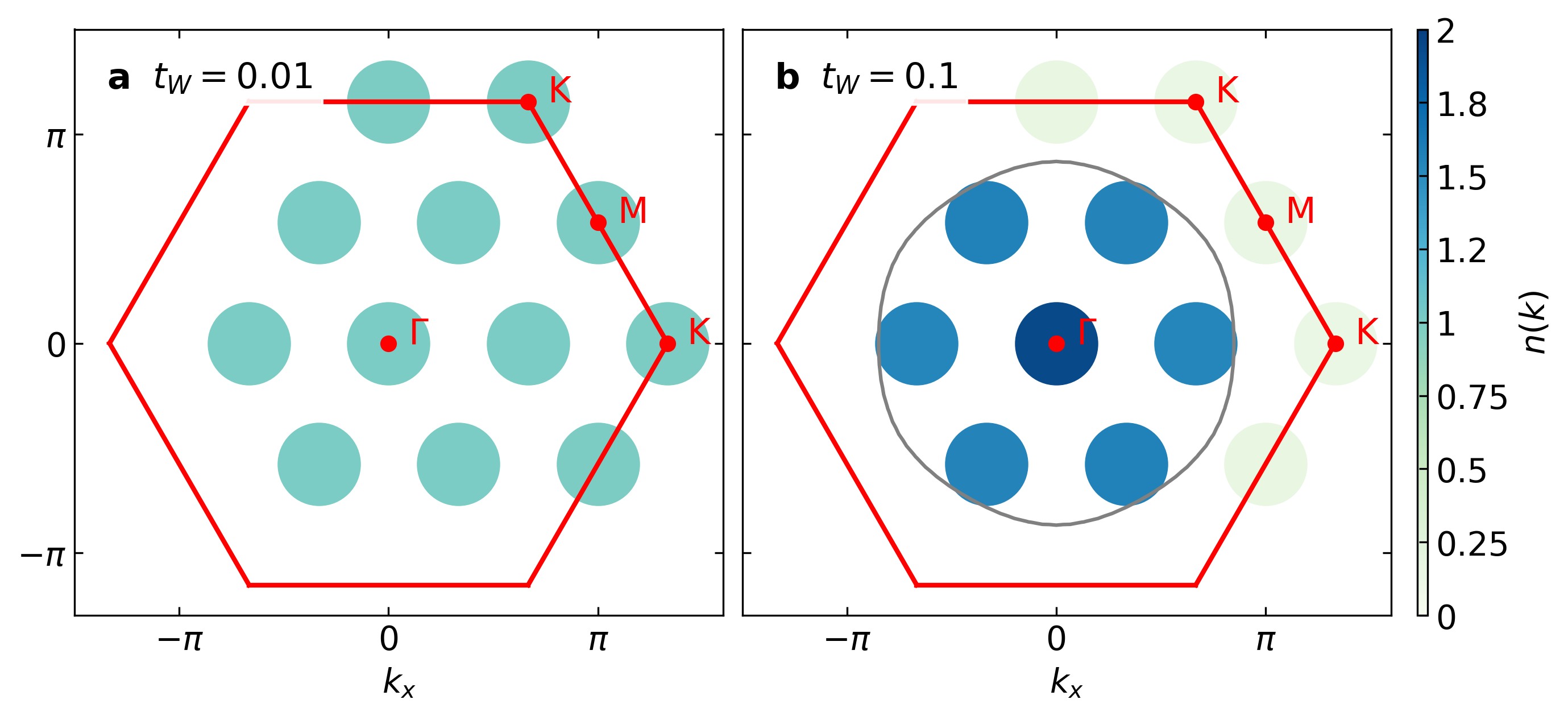}
    \caption{Two representative momentum distribution functions, $n(\vec{k})$, of the Kane-Mele model, at fixed $M=0.6$, $U=1$. \textbf{a} $t_W=0.01$, representative of Mott insulating phase. \textbf{b} $t_W=0.1$, representative of metallic phase. Grey line denotes expected Fermi surface in the thermodynamic limit.}
    \label{fig:nk-sample}
\end{figure}

\begin{figure}[htbp]
    \centering
    \includegraphics[width=0.45\textwidth]{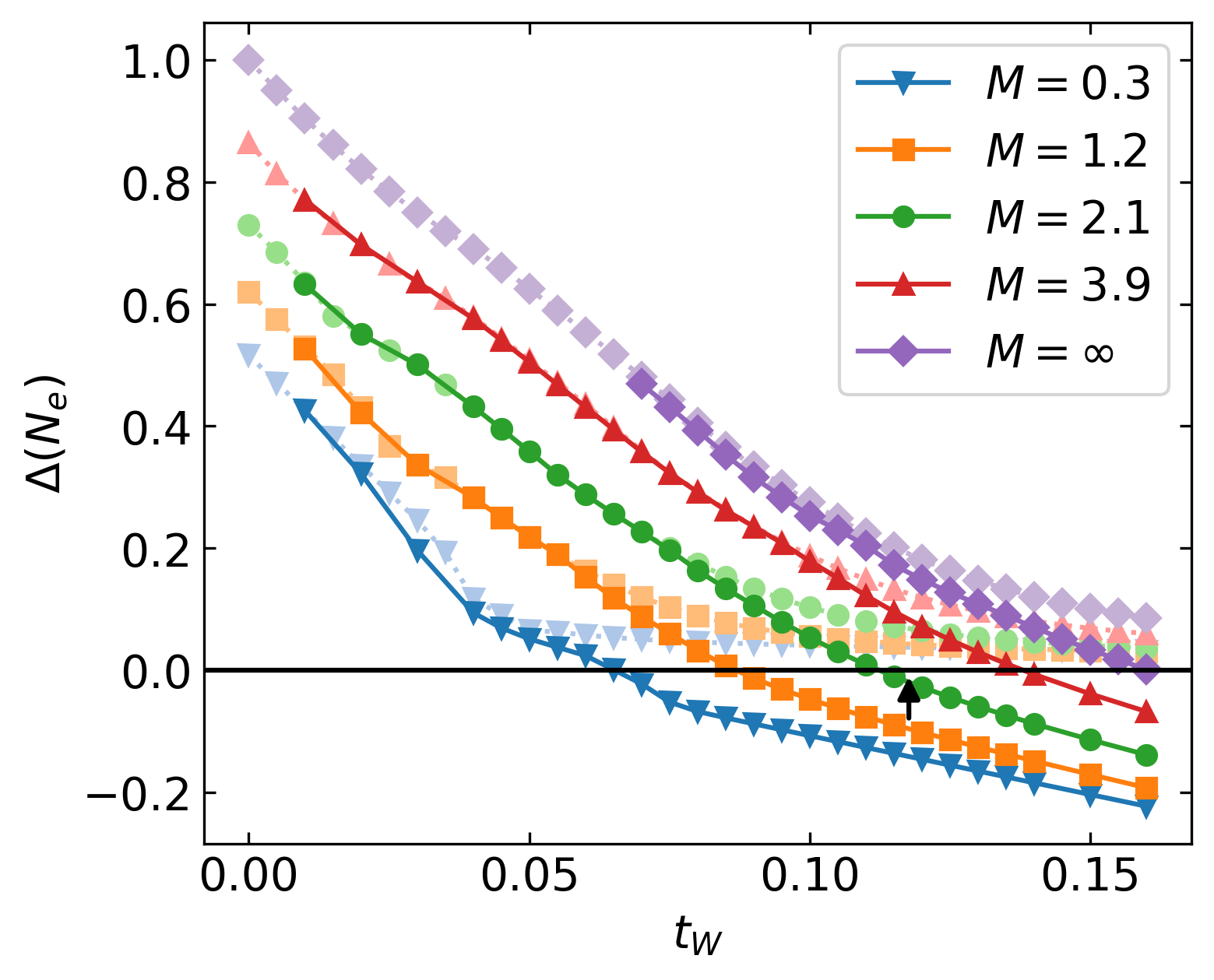}
    \caption{Charge gap obtained by averaging over twisted boundary conditions on a \texttt{12d4} cluster. $M = \infty$ represents the single band triangular lattice Hubbard model. Solid lines represent results obtained after averaging over $4\times 4$ boundary twists, while dotted lines are results obtained with periodic boundary conditions (i.e. without boundary twists, same as \cref{fig:km-phase}\textbf{b}). Black arrow denotes estimate of MIT location in the triangular lattice Hubbard model based on DMRG~\cite{Szasz2020}.}
    \label{fig:gap-vs-tW-KM}
\end{figure}

\begin{figure}[htbp]
    \centering
    \includegraphics[width=0.6\textwidth]{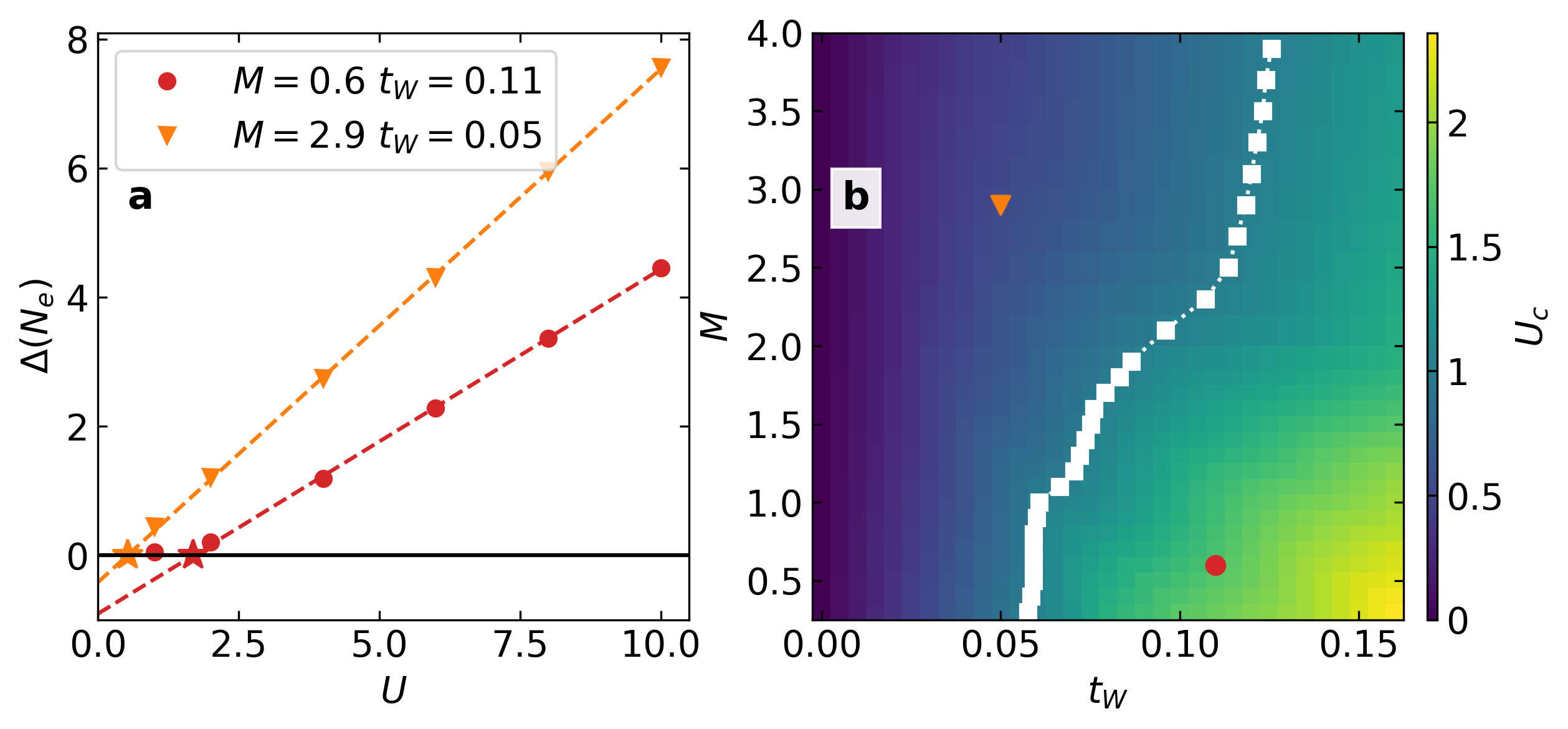}
    \caption{Fit procedure for estimating the metal-to-insulator transition point, $U_c$, in the Kane-Mele model. \textbf{a} Representative linear fits to $\Delta(N_e,U) = aU-b$ for two parameter sets. The parameters choices are also annotated in \textbf{b} with corresponding markers. The range of $U$ values used in the fit is $U \in \{2,4,6,8,10\}$. \textbf{b} Color map of $U_c$ obtained via the $x$-intercept $U_c = -b/a$. White squares denote where $U_c = 1$.}
    \label{fig:Uc-fit}
\end{figure}

\begin{figure}[htbp]
    \centering
    \includegraphics[width=0.4\textwidth]{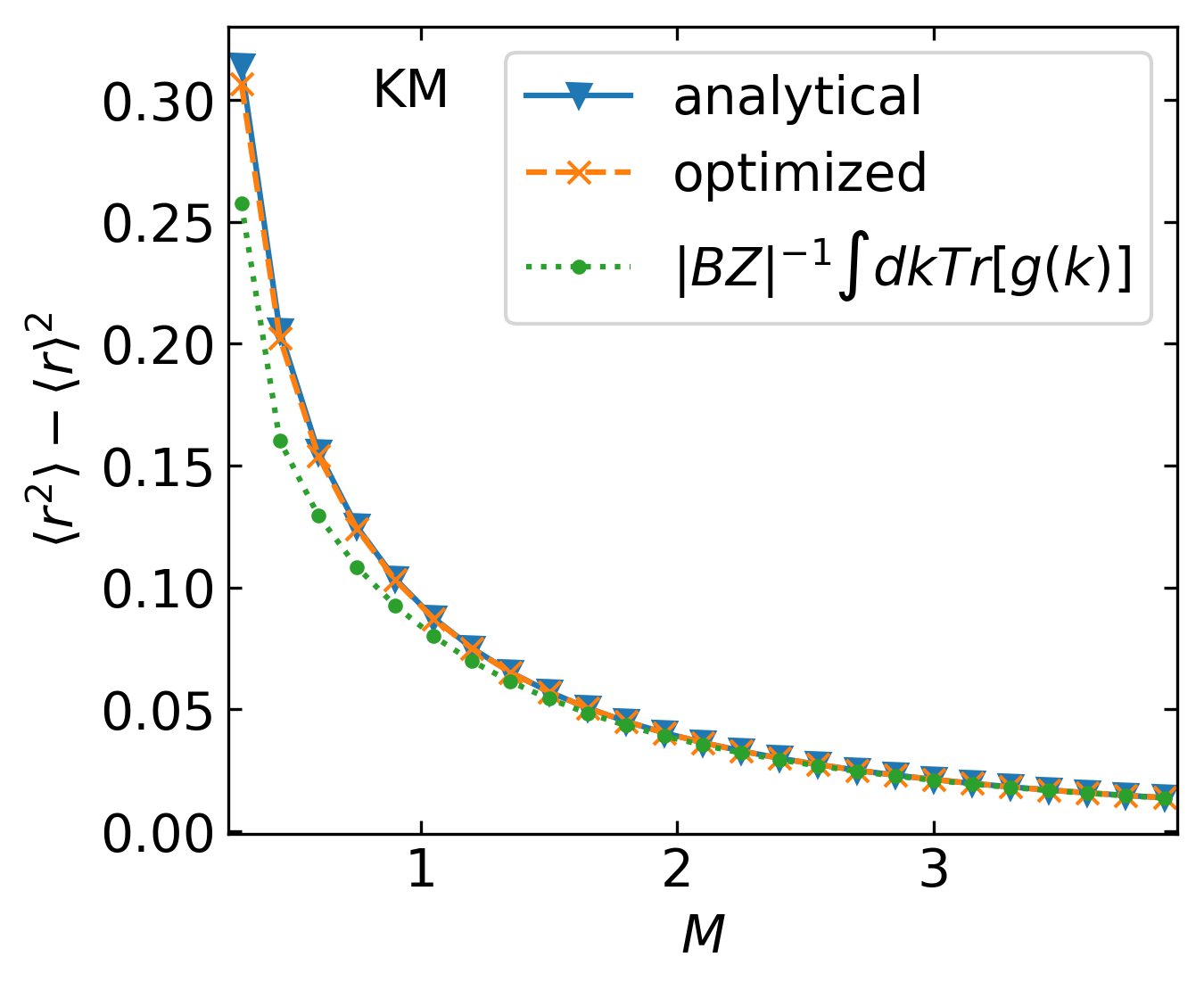}
    \caption{Comparison of analytical gauge choice used in the main text for the Kane-Mele model (\cref{eq:2b-uB}) vs. maximally-localized Wannier functions~\cite{Marzari2012}. }
    \label{fig:gauge-choice-KM}
\end{figure}

\clearpage
\section{Additional results in the BHZ-Hubbard Model}
\label{sec:additional_results_in_the_bhz_hubbard_model}
To make closer contact with the standard square-lattice Hubbard model, we also study the BHZ model, which may be viewed as augmenting the single-band Hubbard model with an extra orbital and introducing quantum geometry by allowing some orbital mixing. The BHZ model consists of two conjugate copies of the QWZ model~\cite{Qi2006} on a square lattice, with
\begin{equation}
\vec{a}(\vec{k}) = \begin{bmatrix}
        t_1 \sin k_x, & t_1 \sin k_y, & M - t_2(\cos k_x + \cos k_y)
    \end{bmatrix}^T. \label{eq:qwz-params}
\end{equation} 
After fixing $t_1 = t_2 = 1$, 
The BHZ model undergoes topological phase transitions at $M=-2,0,2$, at which the spin Chern number of the lower band changes as $C_s = 0 \rightarrow -2 \rightarrow 2 \rightarrow 0$.
We focus on the topologically trivial case of $M>2$ and solve \cref{eq:proj-final-momentum} with added dispersion
\begin{equation}
    \varepsilon(\vec{k}) = -2t_W \left(\cos k_x + \cos k_y\right). \label{eq:sq-dispersion}
\end{equation}
Analogous to the Kane-Mele model, $M$ controls the quantum geometry of this model, and smaller $M$ corresponds to larger orbital mixing and more delocalized Wannnier functions.

\begin{figure}[htbp]
    \centering
    \includegraphics[width=0.65\textwidth]{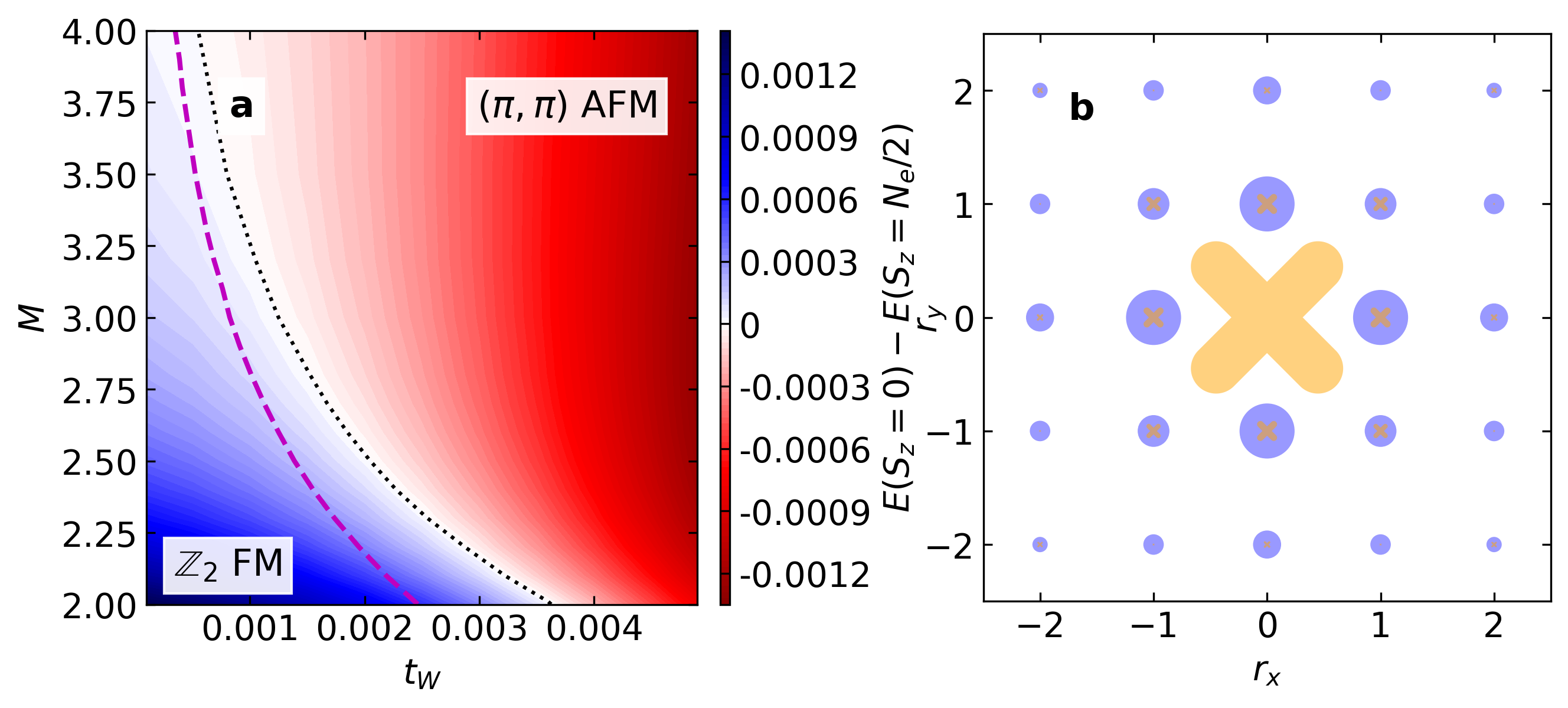}
    \caption{Quarter-filled BHZ model results. \textbf{a} Magnetic phase diagram. Magenta dashed line denotes where nearest neighbor exchange $J^{zz}_1$ changes sign in the projected interaction Hamiltonian, written in Wannier basis. Results are obtained on a square \texttt{10h3} Betts cluster~\cite{Betts1996,Betts1999,Gustiani2015}. \textbf{b} Representative real space spread of the Wannier function of the lower band of the BHZ model at $M=2.1$. The size of the markers is proportional to the amplitude $\abs*{W_{\vec{R}=\mathbf{0}}(\vec{r})}$ on the original orbitals of the square lattice, with purple discs for orbital $A$ and orange crosses for orbital $B$.} 
    \label{fig:qwz-phase}
\end{figure}

\begin{figure}[htbp]
    \centering
    \includegraphics[width=0.5\textwidth]{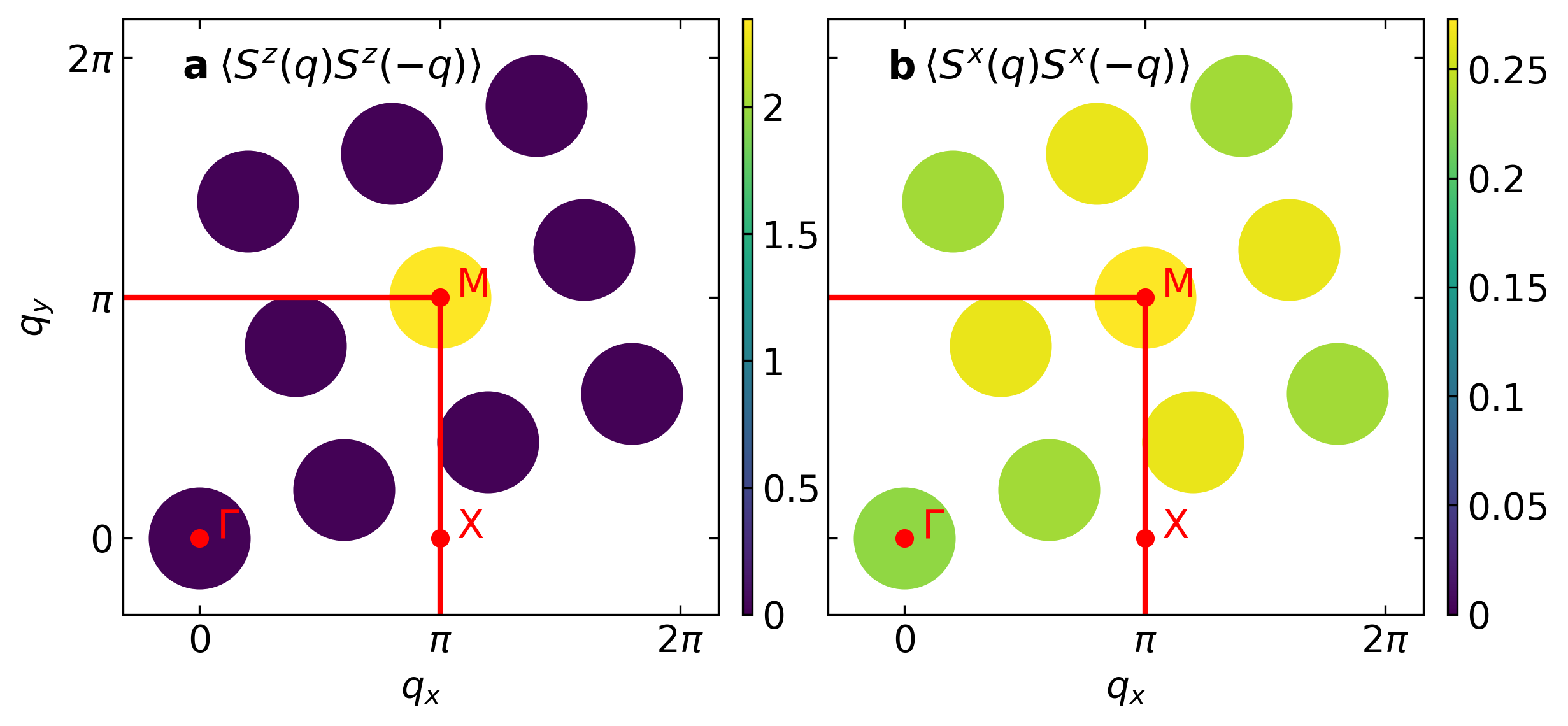}
    \caption{Band-projected spin correlation functions of BHZ model in $(\pi,\pi)$ AFM phase on \texttt{10h3} cluster, demonstrating out-of-plane order at $(\pi,\pi)$. Parameters: $M=2.7$, $t_W = 0.0025$.}
    \label{fig:qwz-spincorr-10h3}
\end{figure}

\begin{figure}[htbp]
    \centering
    \includegraphics[width=0.7\textwidth]{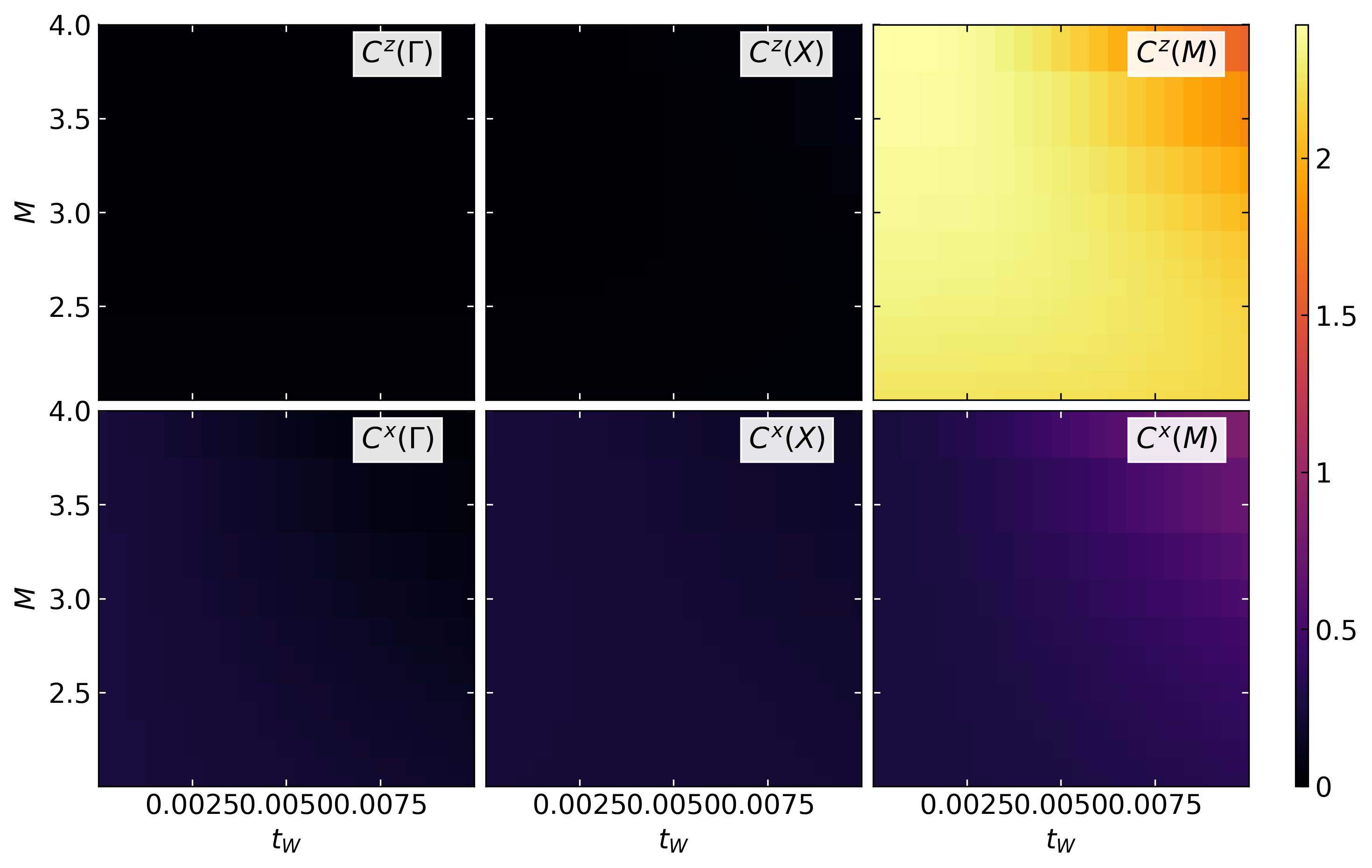}
    \caption{Band-projected spin correlation functions of the BHZ model in the $S_z = 0$ sector at high symmetry points of the \texttt{10h3} cluster.}
    \label{fig:qwz-high-sym-10h3}
\end{figure}

The ground state phase diagram for the BHZ model obtained on a square \texttt{10h3} Betts cluster is shown in \cref{fig:qwz-phase}\textbf{a}. In contrast to the Kane-Mele model, we find that for the range of $t_W$ we study, the ground state is always a Mott insulator. As such, only the spin order is shown. Similar to the Kane-Mele model, the BHZ model transitions between Mott FM and Mott AFM ground states as $M$ and $t_W$ is varied. At small $t_W$, the ground state has out-of-plane $\mathbb{Z}_2$ FM order, identified by the maximal $S_z = N_e/2$ sector having the lowest energy. Whenever the $S_z=0$ sector has the lowest energy, we find that the band-projected spin correlation functions are strongly anisotropic, with the the out-of-plane spin correlation function $C^z(\vec{q})$ having a large peak at $\vec{q} = (\pi,\pi)$ (shown in \cref{fig:qwz-spincorr-10h3,fig:qwz-high-sym-10h3}).

\begin{figure}[htbp]
    \centering
    \includegraphics[width=0.4\textwidth]{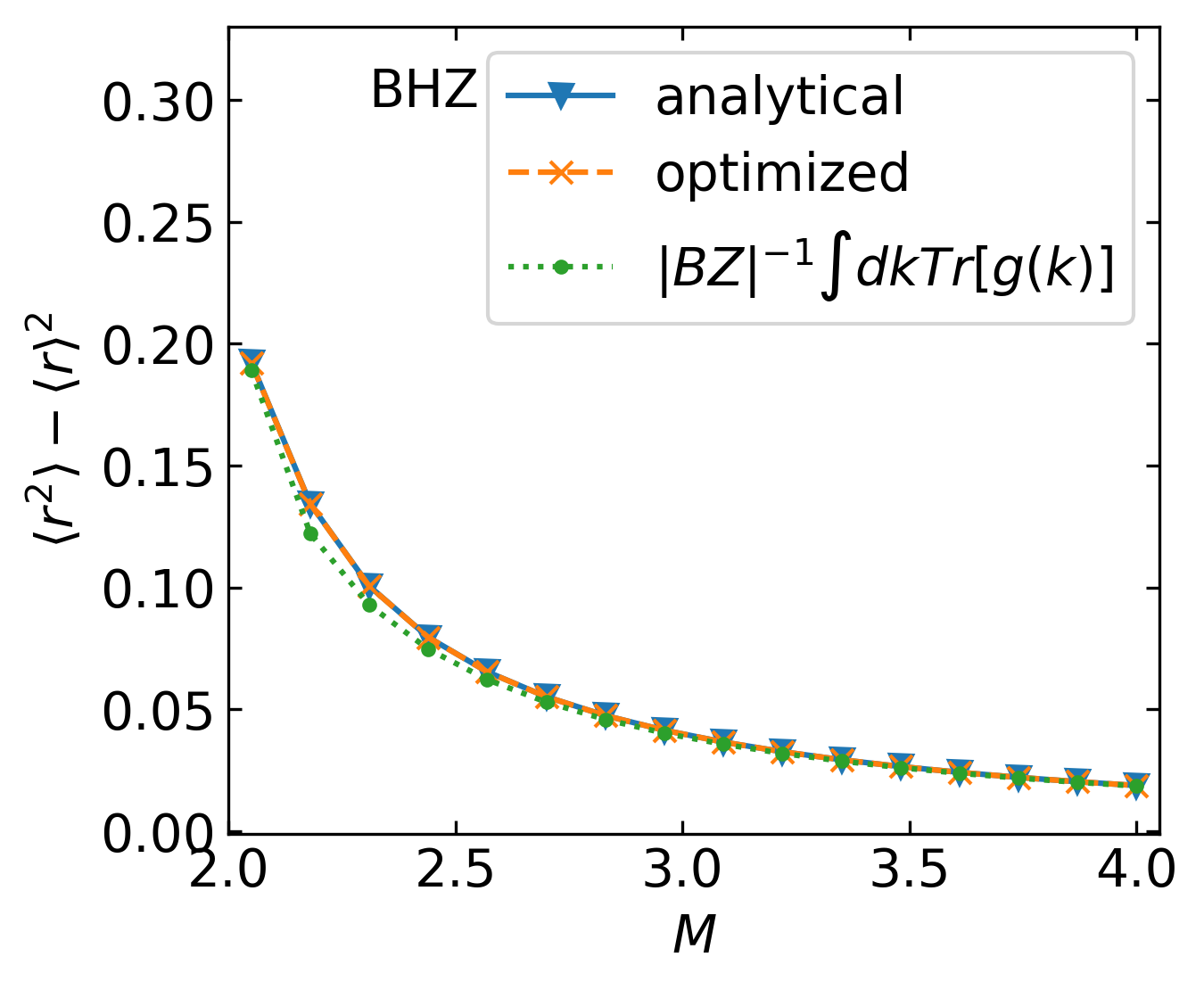}
    \caption{Comparison of analytical gauge choice used for the BHZ model (\cref{eq:2b-uB-qwz}) vs. maximally-localized Wannier functions~\cite{Marzari2012}. }
    \label{fig:gauge-choice-QWZ}
\end{figure}

\begin{figure}[htbp]
    \centering
    \includegraphics[width=0.4\textwidth]{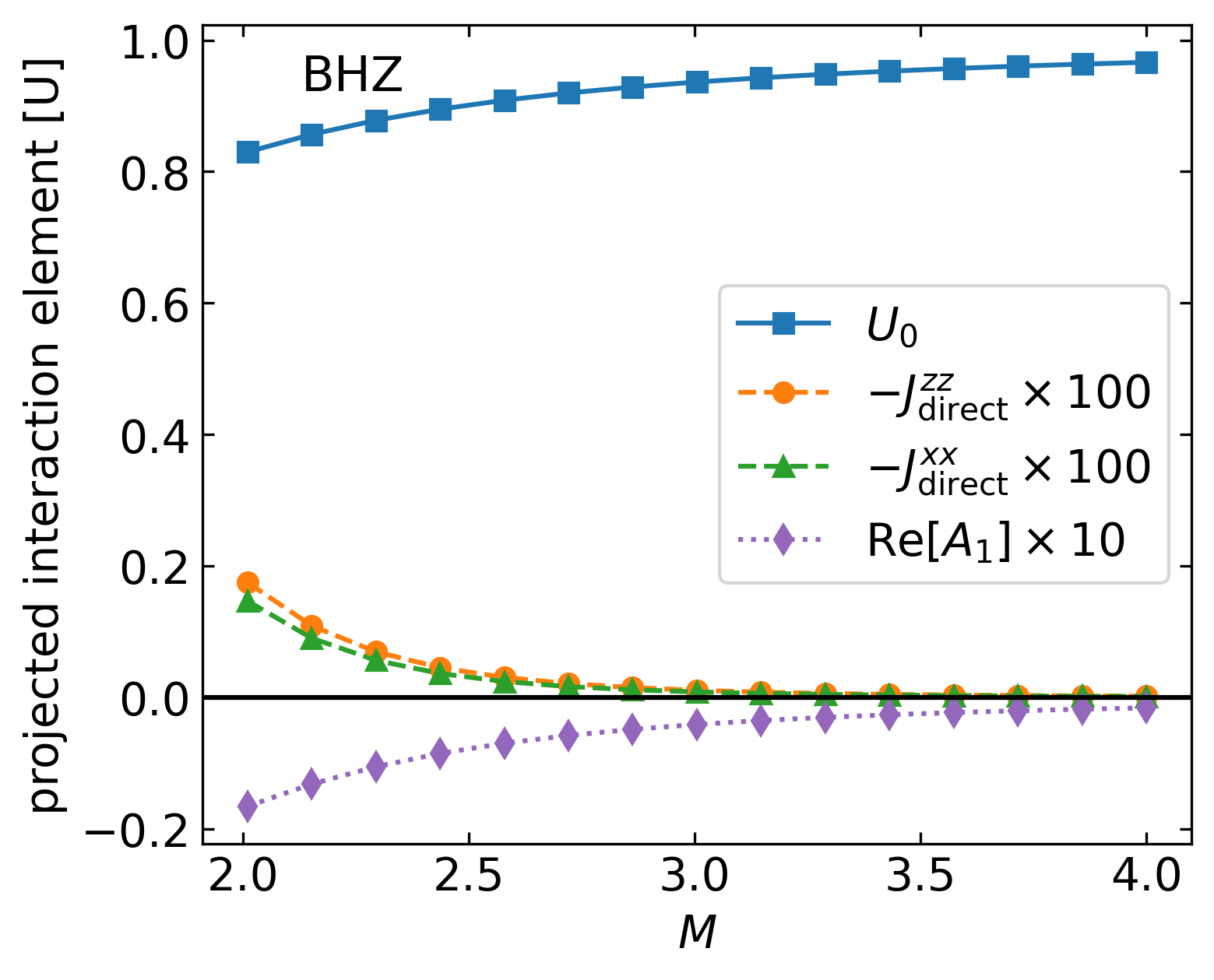}
    \caption{Nearest-neighbor projected interaction elements for Wannier orbitals displaced by $\vec{a}_1$ for the BHZ model, analogous to \cref{fig:int-short-KM} in the main text. In the BHZ model, $\Gamma^{xy}_{\mathrm{direct}}$ and $\mathrm{Im}[A_1]$ are not shown because they are uniformly zero. Except $U_0$, all other interaction terms have been re-scaled to improve visibility. Effective nearest-neighbor repulsion and pair hopping are related to spin exchange matrix elements via $U_1 = - J^{zz}_{\mathrm{direct}}/4$, and $P_1 = -J^{zz}_{\mathrm{direct}}/2$.}
    \label{fig:int-short-QWZ}
\end{figure}

To understand the magnetic transition, we follow the procedure outlined in \cref{eq:V-wannier} through \cref{eq:J-super} and consider short-range interaction elements in a basis of exponentially localized Wannier functions.  Adapting the gauge choice
\begin{equation}
\ket{u^\uparrow_{n=0}(\vec{k})} =
\frac{1}{\sqrt{2|\vec{a}| (|\vec{a}|+a_3)}} \begin{bmatrix}
ia_2-a_1\\
a_3 + |\vec{a}|
\end{bmatrix}, \label{eq:2b-uB-qwz}
\end{equation}
produces almost maximally-localized Wannier functions, as shown in \cref{fig:gauge-choice-QWZ}. A typical Wannier function for the BHZ model is shown in \cref{fig:qwz-phase}\textbf{b}. Using these Wannier functions, an analytical estimate of the location of the magnetic phase transition using the sign of the nearest neighbor exchange $J_1^{zz}(M,t_W)$ (see \cref{fig:int-short-QWZ} for details) is shown as a magenta dashed line in \cref{fig:qwz-phase}\textbf{a}, which agrees reasonably well with numerical results. The FM-AFM transitions observed in the BHZ and Kane-Mele model are qualitatively similar, indicating that the ability for quantum geometry to tune between different magnetic Mott insulating phases are generic. 

\begin{figure}[htbp]
    \centering
    \includegraphics[width=0.4\textwidth]{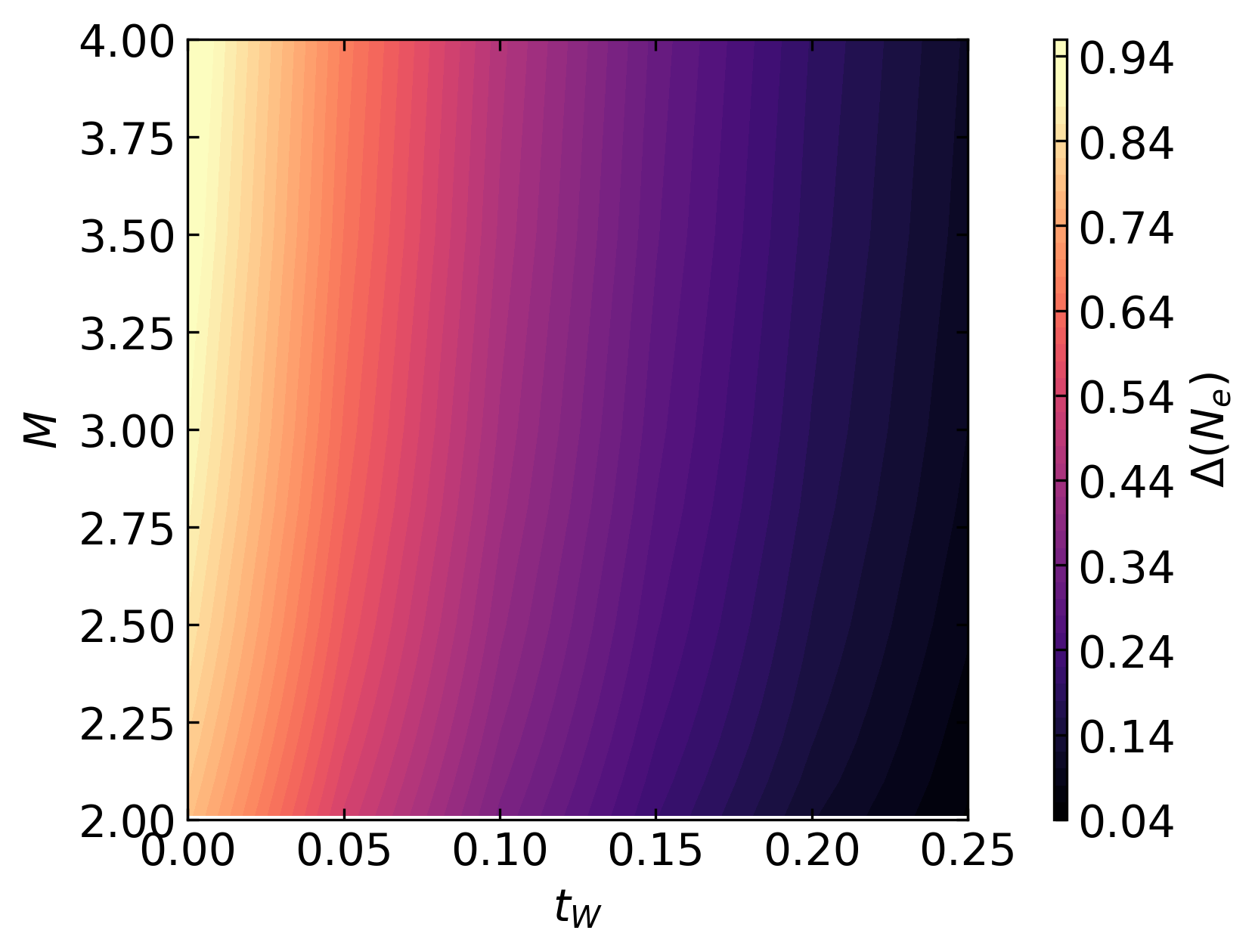}
    \caption{Ground state charge gap of the BHZ model obtained on square \texttt{14h4} cluster. The single-particle dispersion is set as $\varepsilon(\vec{k}) =  -2t_W (\cos k_x + \cos k_y) + -4 t_W' \cos k_x \cos k_y$ with $t_W'/t_W = -0.25$ in order to remove exact Fermi surface nesting (which occurs when $t_W'/t_W = 0$).}
    \label{fig:qwz-14h4}
\end{figure}

The lack of a metallic phase in the range of parameters we studied can be partially attributed to the difference in lattice geometry. The single-particle dispersion on a square lattice \cref{eq:sq-dispersion} at half filling has exact $(\pi,\pi)$ nesting. In the absence of quantum geometric effects (equivalently, in the $M\rightarrow \infty$ limit), the single-band Hubbard model is an insulator at half filling for any nonzero $U$. 
In the BHZ model, next-nearest neighbor correlated hopping induced by quantum geometrical effect weakly breaks this exact Fermi-surface nesting, but this effect is small due to the exponential decay of Wannier functions, making it hard to observe via ED. Nevertheless, we expect the quantum-geometric Mott transition is a generic phenomenon, and not merely limited to the Kane-Mele model. 
Breaking exact Fermi surface nesting by including next-nearest neighbor hopping may enable observation of this transition in square-lattice multiorbital Hubbard models, as shown in \cref{fig:qwz-14h4}.

\end{document}